\documentclass[superscriptaddress,floatfix,twocolumn,amsmath,amssymb]{revtex4-2}
\usepackage{graphicx}
\usepackage{hyperref}
\usepackage{amsmath}
\usepackage{amssymb}
\usepackage{amsthm}
\usepackage{color}
\usepackage{footnote}
\usepackage[normalem]{ulem}
\usepackage{dcolumn}
\usepackage{ragged2e}
\usepackage{multirow}
\usepackage{xcolor}
\usepackage{flushend}
\usepackage{microtype}
\usepackage{siunitx}
\usepackage{caption}
\usepackage{soul}
\usepackage{cancel}
\usepackage{algorithm}
\usepackage{algpseudocode}
\usepackage{braket}
\UseRawInputEncoding
\usepackage{enumitem}
\DeclareMathOperator{\Tr}{Tr}

\DeclareMathOperator{\sgn}{sgn}
\newtheorem{theorem}{Theorem}

\begin{abstract}
State preparation is a cornerstone of quantum technologies, underpinning applications in computation, communication, and sensing. Its importance becomes even more pronounced in non-Markovian open quantum systems, where environmental memory and model uncertainties pose significant challenges to achieving high-fidelity control. Invariant-based inverse engineering provides a principled framework for synthesizing analytic control fields, yet existing parameterizations often lead to experimentally infeasible, singular pulses and are limited to simplified noise models such as those of Lindblad form. 
Here, we introduce a generalized invariant-based protocol for finite-dimensional state preparation under arbitrary noise conditions. We transform the finite-dimensional control problem into the equivalent problem for a single-qubit, by restricting the dynamics to a designed SU(2) subspace. The control protocol then proceeds in two-stages: first, we construct a family of bounded pulses that achieve perfect state preparation in a closed system; second, we identify the optimal member of this family that minimizes the effect of noise. The framework accommodates both (i) characterized noise, enabling noise-aware control synthesis, and (ii) uncharacterized noise, where a noise-agnostic variant preserves robustness without requiring a master-equation description. 
Numerical simulations demonstrate high-fidelity state preparation across diverse targets while producing smooth, hardware-feasible control fields. This singularity-free framework extends invariant-based control to realistic open-system regimes, providing a versatile route toward robust quantum state engineering on NISQ hardware and other platforms exhibiting non-Markovian dynamics.
\end{abstract}

\begin{document}
\title{Singularity-free dynamical invariants-based quantum control}

\author{Ritik Sareen}
\email{s4068992@student.rmit.edu.au}
\address{Quantum Photonics Laboratory and Centre for Quantum Computation and Communication Technology, RMIT University, Melbourne, VIC 3000, Australia}
\address{School of Electrical Engineering and Telecommunications, UNSW Sydney, Sydney, NSW 2052, Australia}

\author{Akram Youssry}
\address{Quantum Photonics Laboratory and Centre for Quantum Computation and Communication Technology, RMIT University, Melbourne, VIC 3000, Australia}
\address{School of Electrical Engineering and Telecommunications, UNSW Sydney, Sydney, NSW 2052, Australia}

\author{Alberto Peruzzo}
\address{Quantum Photonics Laboratory and Centre for Quantum Computation and Communication Technology, RMIT University, Melbourne, VIC 3000, Australia}
\address{Quandela, Massy, France}

\maketitle

\section{Introduction}\label{sec:intro}
Quantum state preparation is a fundamental prerequisite for many applications of quantum technology. The goal is to steer a quantum system from a fixed fiducial initial state to a desired target state-such as a computational basis state or an entangled state. Achieving precise state preparation underpins advances in quantum computing~\cite{qtqcomp,qspalg}, communication~\cite{qtqcomm}, error correction~\cite{qspqec}, sensing~\cite{qtqsens}, and metrology~\cite{qtqmet}. However, engineered quantum systems are intrinsically susceptible to noise and decoherence, making accurate and robust control a central challenge.

A variety of control strategies have been developed to address this problem, including quantum control landscapes \cite{rabitz_landscape_2004, riviello_constraints_2015}, Gradient Ascent Pulse Engineering (GRAPE)~\cite{grape}, Chopped Random-Basis (CRAB)~\cite{crab1,crab2}, variational quantum algorithms~\cite{vqe,qaoa}, feedback-based control~\cite{feedqsp1,feedqsp2}, reinforcement learning approaches~\cite{qsprl1,qsprl2,qsprl3}, and adiabatic techniques~\cite{adia1,adia2,adia3}. Among these, adiabatic methods have proven particularly powerful for implementing robust quantum gates~\cite{robadia1,robadia2,robadia3} and for analog quantum optimization~\cite{aqc,qann}, offering an alternative to digital quantum algorithms such as Grover's search~\cite{grover1996fast}, the Quantum Approximate Optimization Algorithm (QAOA)~\cite{qaoa}, and the Variational Quantum Eigensolver (VQE)~\cite{vqe}. While Grover’s algorithm requires fault-tolerant quantum hardware, QAOA and VQE may fail to find global optima or deliver genuine quantum advantage. A major limitation of adiabatic control, however, lies in its inherently long evolution times, which make it vulnerable to decoherence. This challenge has motivated the development of fast control protocols collectively known as shortcuts to adiabaticity (STA)~\cite{sta1,sta2}.

One prominent family of STA techniques relies on Lewis-Riesenfeld invariants~\cite{lewis1969exact}, also referred to as dynamical invariants. Controlling the system Hamiltonian through these invariants is often referred to as invariant-based inverse engineering~\cite{inveng1,inveng2,inveng3,inveng4}. A common formulation employs Lie algebraic methods~\cite{invlie1,invlie2,invlie3,invlie4,invlie5,invlie6}, where invariant-based pulse engineering  has been proposed for closed quantum systems of various dimensionalities. More recently, this framework has been extended to state preparation in Markovian open systems~\cite{levy2018noise,invopen1} and to the implementation of quantum gates for cat-state photonic qubits~\cite{invcat}. However, the parametrizations used in most of these studies can give rise to singular control pulses with one or more points diverging to infinity \cite{singular,sing1,sing2} as hinted in \cite{confession}, limiting their physical feasibility. Consequently, such pulse design methods are typically restricted to specific pairs of initial and final states and cannot be straightforwardly generalized. Moreover, existing invariant-based approaches have been applied primarily to closed systems or those subject to Markovian noise.

Non-Markovian dynamics~\cite{nmqd}, which frequently occur in realistic quantum devices~\cite{nmexp1,nmexp2}, remain particularly challenging to model and control. In these systems, memory effects in the environment feed back into the system dynamics, making them difficult to characterize without detailed environmental information. As a result, a Lindblad-form master equation with time-independent collapse operators~\cite{gksl1,gksl2} cannot describe such noise processes. Alternative formulations, such as the Redfield equation~\cite{redfield1957theory} and the Nakajima–Zwanzig (NZ) equation~\cite{nze1,nze2}, have been proposed. The Redfield equation, however, is valid only in the weak-coupling or Markovian limit and can even yield non-physical results, whereas the NZ equation, though capable of capturing non-Markovian effects, is often computationally intractable. The approach in Ref.~\cite{invopen1} demonstrated state preparation for systems described by time-dependent Lindblad master equations, applicable only to limited noise classes-for instance, not to classical colored noise.

In this paper, we propose a general framework for invariant-based quantum state preparation in non-Markovian open quantum systems that overcomes these limitations. Our first contribution is a protocol that generates a family of control pulses capable of steering a closed two-level system from an arbitrary initial to an arbitrary final state. Second, we rigorously prove that all pulses produced by our method are guaranteed to bounded, i.e. remain finite in amplitude at all times and therefore singularity-free. Next, we extend invariant-based control beyond Lindblad-type dynamics by incorporating an optimal control layer that selects, from the family of feasible pulses, the one that minimizes the impact of non-Markovian noise on system evolution. We consider two distinct settings: (i) when the noise characteristics are known, we employ an analytical cost function expressed through perturbative expansions, and (ii) when the noise is unknown, we embed a model-based supervised machine-learning (ML) module into the control loop. Numerical simulations confirm that our protocol achieves high-fidelity state preparation for arbitrary qubit targets under multi-axis classical colored noise.

Furthermore, we introduce an approach to extend the proposed method to finite-dimensional quantum systems assuming full control with respect to a complete set of basis. By restricting the control dynamics to an SU(2) subspace, the complexity of the control problem can be drastically reduced. This approach is inspired by several ideas that appear in different contexts in the quantum literature. This includes, for example, the concept of logical qubits in quantum error correction \cite{qec1,qec2,qspqec} and dressed states in light-matter interaction  \cite{dressed1,dressed2}. While error correction protocols are based on encoding qubits into higher-dimensional systems to enable detection and correction of errors; here, we utilize a smaller SU(2) subspace in the qudit space. A similar approach is also observed in cat-encoded qubits \cite{cat1,cat2}, which are utilized in many error correcting protocols and fault-tolerant quantum computation \cite{catapp1,catapp2,catapp3}. Recently, STA methods have been used to prepare and control cat qubits \cite{catsta,invcat}. In the context of quantum optics, dressed states are often used to define an effective qubit out of a higher-dimensional system such as a two-level atom interacting with a quantum field. The concept of designing control restricted to a subspace has been previously utilized in invariant-based higher-dimensional state preparation \cite{invlie2,invlie4,invlie5}, as well as control of continuous-variable systems \cite{levy2018noise,iontrap,iontrap2,invcat}. Those works assume that the subspace is fixed as a result of some experimental constraints, and thus the methods are customized to a particular system setting and dimensionality. Here we provide a generic approach that works for finite-dimensional systems of arbitrary dimensions. 

Our work is also related to the inverse quantum control and tracking-control literature~\cite{gross1993inverse,zhu1999managing,magann2018singularityfree} but differs in both formulation and scope. In those approaches, the goal is to invert the dynamical equations to enforce the expectation values of selected observables to follow prescribed tracks. This results in the appearance of singular pulses, which is addressed by incorporating higher-order Heisenberg equations of motion. This solution is not suited for the invariant-based approach as the mathematical formulation of the two approaches is different, i.e. different cause for the singularity. In our work, we focus on state perpetration and overcome the singularity by engineering the closed-system trajectory.

This work fills a critical gap in the literature on invariant-based inverse-engineering control. The proposed protocol efficiently yields a family of physically valid control pulses for preparing the ground state of a target Hamiltonian, while allowing the choice of optimality criteria such as minimal energy, robustness, or hybrid performance metrics. The proposed strategy for subspace control of finite-dimensional systems can also be extended to other control methods or specialized to specific quantum platforms and target states. The framework thus represents an important step toward realizing high-performance quantum operations on NISQ devices and opens the door to analog quantum optimization on noisy platforms operating at short time scales.

\section{Problem setting}\label{sec:prob}
In this paper, we address the problem of controlling a finite-dimensional system undergoing general non-Markovian dynamics in order to prepare a desired target state $\ket{\psi_T}$ at time $t=T$, given an initial state $\ket{\psi(0)}$. The total Hamiltonian of the system is given by
\begin{equation}
    H(t) = H_{\text{ctrl}}(t) + H_{\text{noise}}(t),
\end{equation}
where $H_{\text{ctrl}}(t)$ is the control Hamiltonian and $H_{\text{noise}}(t)$ represents the noise affecting the system. The control Hamiltonian for a qudit can be expressed in the general form
\begin{align}\label{quditctrl}
    H_{\text{ctrl}}(t)=\sum_{j=0}^{d^2-1}{c}_j(t)\lambda_j,
\end{align}
where $c_{j}(t)$ denotes the control pulse amplitude for the $j^{\text{th}}$ basis, and $\{\lambda_j\}$ is a complete set of Hermitian orthonormal basis with $\lambda_0$ being the normalized identity matrix $I/\sqrt{d}$, (i.e. we assume we have full control of all qudit bases). The control problem of preparing a target state in any finite-dimensional closed system can be reduced to an equivalent state-preparation problem in a two-level system. Given the initial and target states, we construct the SU(2) subspace spanned by the initial state, and the component of the target state orthogonal to it. By introducing equivalent Pauli operators on this subspace, the pulse-design problem can be formulated in the same way as for a single qubit. Once the required control is obtained in the SU(2) picture, it can be mapped back to the original finite-dimensional system. We show this equivalence and how to transition between qudit and SU(2) picture in Section \ref{sec:equivalence}. Therefore, we will focus on the invariants-based engineering for a two-level system.

The two-level system control Hamiltonian can be expressed generally in the form:
\begin{equation}\label{hctrl}
    H_{\text{ctrl}}(t) = \frac{h_x(t)}{2}\sigma_x + \frac{h_y(t)}{2}\sigma_y + \frac{h_z(t)}{2}\sigma_z,
\end{equation}
where $h_{x,y,z}(t)$ denote the control pulse amplitudes and $\sigma_{x,y,z}$ are the Pauli matrices along the $x$, $y$, and $z$ directions, respectively. We assume full three-axis control of the qubit, which can be achieved in several quantum platforms such as superconducting qubits~\cite{3ctrl}. In contrast to the conventional two-axis setting, full control is required here to ensure that the engineered control pulses remain bounded at all times and for arbitrary target states, as discussed in detail in Section~\ref{subsec:traj}.

The noise affecting the system can be classical, quantum, or a combination of both. In the case of classical noise, $H_{\text{noise}}(t)$ can be modeled as a stochastic system-only operator. For quantum noise, $H_{\text{noise}}(t)$ includes both system and bath operators, incorporating the bath free Hamiltonian and the system-bath interaction. If both noise types are present, then $H_{\text{noise}}(t)$ contains both stochastic and quantum bath terms. When quantum noise is present, we assume that the system and bath are initially uncorrelated, i.e. $\rho(0) = \ket{\psi(0)}\!\bra{\psi(0)} \otimes \rho_B$, where $\rho_B$ is the initial state of the bath.

To solve the state-preparation control problem, we decompose it into two subproblems:

\paragraph*{(1) Invariant-based pulse engineering.}
The first subproblem is to design the control pulses that prepare the desired final state in the absence of noise, i.e. $H(t)=H_{\text{ctrl}}(t)$. We require all control amplitudes to remain bounded, $h_x(t), h_y(t), h_z(t) < \infty,\ \forall t \in [0,T]$. The initial state $\ket{\psi(0)}$ is taken to be the ground state of the initial control Hamiltonian $H_{\text{ctrl}}(0) = \tfrac{1}{2}(h_x(0)\sigma_x + h_y(0)\sigma_y + h_z(0)\sigma_z)$, and the target state $\ket{\psi_T}$ is the ground state of the final Hamiltonian $H_{\text{ctrl}}(T) = \tfrac{1}{2}(h_x(T)\sigma_x + h_y(T)\sigma_y + h_z(T)\sigma_z)$. Supplementary Note~1 describes how to determine these boundary Hamiltonians given the two states. The state-preparation problem then reduces to finding $H_{\text{ctrl}}(t)$ for $0 < t < T$ that evolves the ground state of the initial Hamiltonian to that of the final one. This can be solved elegantly using invariant-based pulse engineering. The dynamical-invariant framework yields an infinite family of control pulses that all exactly solve the closed-system state-preparation problem, though their intermediate dynamics differ. This flexibility allows one to select a specific pulse according to an additional optimality criterion central to addressing noise in the next subproblem.

\paragraph*{(2) Noise mitigation.}
Dynamical invariants are defined for closed systems. However, in most practical settings noise is unavoidable, requiring further treatment to extend the approach. The presence of noise breaks the equivalence among the pulse family obtained in subproblem~1: while all pulses yield the same final target state in the absence of noise, each produces a different final state when noise is present. This observation motivates the use of dynamical invariants in open systems—by selecting a pulse from the invariant-based family that minimizes the impact of noise. If a pulse exists that effectively cancels the noise contribution, then by construction the final state converges to the target despite the environment. The existence of such an ``optimal'' pulse is a quantum controllability problem that remains an open question and is beyond the scope of this paper. Instead, we seek a control pulse that minimizes the effect of noise, rendering the final state as close as possible to the target. This concept has been explored previously for systems governed by Lindblad-type master equations~\cite{levy2018noise,invopen1}. Here, we generalize it to encompass non-Markovian dynamics. Specifically, we consider two settings: (i) when the noise model is known a priori, and (ii) when the noise is unknown.

Finally, to assess the performance of the optimal pulse obtained from both subproblems, we compute the fidelity between the final state of the system $\rho(T)$ and the target state $\ket{\psi_T}$, defined as
\begin{equation}\label{eq:fidelity}
    \mathcal{F}(T) = \bra{\psi_T}\rho(T)\ket{\psi_T}.
\end{equation}
where $\rho(T)$ is the generally mixed final state of the system. This definition allows for the realistic case in which noise effects are mitigated but not completely eliminated.

\section{Preliminaries}\label{sec:prelim}
\subsection{Dynamical invariants}
A dynamical invariant $I(t)$ is a Hermitian system operator satisfying the Lewis-Riesenfeld condition~\cite{lewis1969exact},
\begin{equation}\label{eqinvariant}
    \frac{\partial I(t)}{\partial t} + i[H_{\text{ctrl}}(t), I(t)] = 0
\end{equation}
which defines its evolution under the control Hamiltonian $H_{\text{ctrl}}(t)$. A fundamental property of $I(t)$ is that its eigenvalues remain constant in time. If $\ket{\phi_k(t)}$ denotes the $k^\text{th}$ instantaneous eigenstate of $I(t)$, then
\begin{align}
    I(t)\ket{\phi_k(t)}=\mu_k \ket{\phi_k(t)}
\end{align}
where the eigenvalue $\mu_k$ is time independent. Moreover, the eigenstates of the invariant evolve according to the Schr\"odinger equation up to a global phase $\alpha_k(t)$,
\begin{align}
    i \hbar \frac{\partial}{\partial t} \left(e^{i\alpha_k(t)} \ket{\phi_k(t)}\right)= H_{\text{ctrl}}(t)\left(e^{i\alpha_k(t)} \ket{\phi_k(t)}\right).
\end{align}

This relation implies that if the system is initially prepared in the ``ground state'' $\ket{\phi_1(0)}$ of the invariant $I(0)$, corresponding to the smallest eigenvalue $\mu_1$, it will evolve to the ground state $\ket{\phi_1(T)}$ of the final invariant $I(T)$ up to a global phase $\alpha_1(T)$. Specifically, if $\ket{\psi(0)} = \ket{\phi_1(0)}$, the final state is
\begin{align}
    \ket{\psi(T)} &=  U_{\text{ctrl}}(T)\ket{\phi_1(0)} \\
    &= e^{i\alpha_1(T)}\ket{\phi_1(T)},
\end{align}
where 
\begin{equation}\label{uctrl}
    U_{\text{ctrl}}(T)=\mathcal{T}_+e^{-i\int_{0}^{T}H_{\text{ctrl}}(t)ds},
\end{equation}
and the accumulated phase $\alpha(T)$ is given by
\begin{equation}\label{eq:phase}
    \alpha_1(T) = \int_0^T\bra{\phi_1(t)}i\frac{\partial}{\partial t} - H_{\text{ctrl}}(t)\ket{\phi_1(t)}\,dt.
\end{equation}
If the system starts in an eigenstate of $I(0)$, the global phase $\alpha_k(T)$ can be neglected, as it does not affect expectation values of the observable. However, for an initial superposition of eigenstates, relative phases become relevant. For instance, if $I(0)=-\sigma_z$, $I(T)=\sigma_z$, and $\ket{\psi(0)}=\ket{+}$, then
\begin{equation}
    \ket{\psi(T)} = \frac{e^{i\alpha_1(T)}\ket{1} + e^{i\alpha_2(T)}\ket{0}}{\sqrt{2}},
\end{equation}
which highlights the importance of relative phases for gate implementation~\cite{invcat}. In this work, we focus on ground-state-to-ground-state preparation, and thus the phase can be disregarded. Finally, by imposing the boundary condition
\begin{align}
    [I(0), H_{\text{ctrl}}(0)] = [I(T), H_{\text{ctrl}}(T)] = 0,
\end{align}
we ensure that an initial ground state of $H_{\text{ctrl}}(0)$ evolves into the ground state of $H_{\text{ctrl}}(T)$.

\subsection{Invariant-based inverse engineering}
The invariant-based inverse engineering protocol~\cite{inveng1,inveng2,inveng3,inveng4,invcat} provides a constructive means of designing control pulses that realize a desired target state or operation. Instead of deriving the invariant from a given Hamiltonian, one begins by postulating an invariant that satisfies prescribed boundary conditions and then infers the corresponding Hamiltonian. This procedure involves defining appropriate constraints on $I(t)$, determining a functional form that meets those constraints, and computing the control fields from Equation \ref{eqinvariant}. In general, infinitely many invariants satisfy the boundary conditions; suitable parameterization therefore yields a continuous family of invariants, and consequently, a family of equivalent control pulses that all achieve perfect state transfer in the closed-system setting.

For a single qubit, the invariant can be expressed as 
\begin{equation}
    I(t) = \frac{f_x(t)}{2}\sigma_x + \frac{f_y(t)}{2}\sigma_y + \frac{f_z(t)}{2}\sigma_z,
\end{equation}
where $f_{x,y,z}(t)$ are coefficients of the invariant along the $X$, $Y$, and $Z$ directions. Substituting this form and the control Hamiltonian of Equation \ref{hctrl} into Equation \ref{eqinvariant} gives
\begin{align}\label{fhrelation}
    \dot{f}_x(t) &= f_z(t)h_y(t) - f_y(t)h_z(t)\nonumber\\
    \dot{f}_y(t) &= f_x(t)h_z(t) - f_z(t)h_x(t)\nonumber\\
    \dot{f}_z(t) &= f_y(t)h_x(t) - f_x(t)h_y(t).
\end{align}
These relations are linearly dependent and admit a solution~\cite{levy2018noise} only if
\begin{align}\label{eq:sphere}
    f_x(t)\dot{f}_x(t) + f_y(t)\dot{f}_y(t) + f_z(t)\dot{f}_z(t) = 0,
\end{align}
which implies that
\begin{align}
    f_x^2(t) + f_y^2(t) + f_z^2(t) = c^2,
\end{align}
representing the conservation of the invariant vector's magnitude. The boundary condition $[I(t_b), H(t_b)] = 0$, for $t_b = 0,T$ leads to
\begin{equation}
    \frac{\partial I(t)}{\partial t} \bigg|_{t=t_b} = 0,
\end{equation}
or equivalently,
\begin{equation}\label{eq:bc1}
    \dot{f}_x(t_b) = \dot{f}_y(t_b) = \dot{f}_z(t_b) = 0.
\end{equation}
Substituting Equation \ref{eq:bc1} into Equation \ref{fhrelation}, results in the following additional boundary conditions:
\begin{align} \label{ihrelation}
    f_z(t_b)h_y(t_b) - f_y(t_b)h_z(t_b) &= 0\nonumber \\
    f_x(t_b)h_z(t_b) - f_z(t_b)h_x(t_b) &= 0\nonumber \\
    f_y(t_b)h_x(t_b) - f_x(t_b)h_y(t_b) &= 0.
\end{align}
At $t=t_b$, the control Hamiltonian is known, so the boundary values of the invariant can be uniquely determined. Any triple of functions $f_{x,y,z}(t)$ satisfying Equations \ref{eq:sphere}, \ref{eq:bc1}, and \ref{ihrelation} solves the control problem. Because of Equation \ref{eq:sphere}, only $f_x(t)$ and $f_y(t)$ need to be specified, while $f_z(t)$ follows from the normalization constraint, reducing the problem’s dimensionality. Finally, by predefining one of the control fields, the remaining two can be computed from Equation \ref{fhrelation}. For example, fixing $h_z(t)$ yields
\begin{align}\label{eq:drift}
    h_x(t) &= \frac{f_x(t)h_z(t) - \dot{f}_y(t)}{f_z(t)}\nonumber\\
    h_y(t) &= \frac{f_y(t)h_z(t) + \dot{f}_x(t)}{f_z(t)}.
\end{align}

This expression requires $f_z(t)\neq 0, \ \forall t \in[0,T]$, a condition that may not hold for certain target states. We address this limitation later in the paper.

\subsection{Modeling non-Markovian dynamics}\label{subsec:model}
Open quantum systems exhibiting non-Markovian dynamics are notoriously difficult to model. Several mathematical approaches have been developed, including cumulant expansions~\cite{cumulant}, filter-function methods~\cite{filter}, time-evolving matrix-product operators~\cite{tempo}, and process tensors~\cite{proten}. In this work, we employ the noise-operator formalism~\cite{vo1,W3,wise}, which provides a compact and general representation of system-environment interactions. In this formulation, the expectation value of a system observable $O$ at time $t=T$ is expressed as
\begin{align}
    \braket{O(T)} = \braket{\Tr_B[U(T)\left(\rho_S(0)\otimes\rho_B\right)U^\dagger(T) (O\otimes \mathbb{I}_B)]}_c,
\end{align}
where $\rho_S(0)$ is the initial system state, $U(T)$ is the total system-bath unitary, and $\mathbb{I}_B$ is the bath identity operator. As shown in Supplementary Note~2, employing a modified interaction picture allows this expression to be recast as
\begin{equation}\label{uvo}
	\braket{O(T)} = \Tr[V_O(T)U_{\text{ctrl}}(T)\rho_S(0)U_{\text{ctrl}}^\dagger(T)O].
\end{equation}
Here, the system operator $V_O$ is defined as
\begin{equation}\label{eq:vo}
	V_O(T) = \braket{O^{-1}\text{Tr}_B[(\tilde{U}_I^{\dagger}(T)O\tilde{U}_I(T))\rho_B]}_c
\end{equation}
where $U_{\text{ctrl}}(T)$ is the control unitary from Equation~\ref{uctrl}, and the average $\langle\cdot\rangle_c$ denotes classical averaging over noise realizations. The modified interaction unitary $\tilde{U}_I(T)$ is defined as $\tilde{U}_I(T) = U_{\text{ctrl}}(T)U_I(T)U_{\text{ctrl}}^\dagger(T)$, with $U_I(T)$ corresponding to the interaction Hamiltonian $H_I(t) = U_{\text{ctrl}}^\dagger(t)H_{\text{noise}}(t)U_{\text{ctrl}}(t)$. 

The noise operator $V_O(T)$ compactly encodes the effect of environmental noise on system dynamics. For a closed system, $V_O(T)=\mathbb{I}_S$, and Equation~\ref{uvo} reduces to the standard expression for quantum expectations. Consequently, minimizing the impact of noise amounts to identifying control fields that make $V_O(T)\!\rightarrow\!\mathbb{I}_S$. This formalism is broadly applicable to finite-dimensional systems, regardless of bath dimensionality, and serves as the foundation for our treatment of non-Markovian open-system control.

\section{Methodology}
Given the initial and target states (or equivalently, initial and target control Hamiltonians), the first step in our proposed protocol is to construct an SU(2) subspace for higher-dimensional quantum systems, effectively transforming the problem to single-qubit control.  Next, we introduce additional constraints on the system trajectory between the initial and final states to pass through predesigned intermediate points. This splits the trajectory into smaller subtrajectories, an essential step to ensure that the control pulses remain bounded throughout the evolution time. This is a major issue that has not been addressed in previous literature on dynamical invariants. We refer to this step as designing the system trajectory.

Next, we apply inverse engineering to find the control pulses that achieves each subtrajectory separately. This begins with finding the boundary invariant that commutes with the boundary Hamiltonian (i.e. the initial and final points of the subtrajectory). Afterwards, we introduce a parameterized functional form of the invariant the satisfies the required constraints and boundary conditions. This gives rise to a family of invariants that achieves the desired evolution. The parameters defining any invariant needs to be bounded, in order to avoid nonphysical (complex-valued or singular) control pulses. Once the family of invariants is designed, we can then find the corresponding family of control pulses. This provides a general method to find the required family of control pulses that ensures the finite-dimensional system evolves from any arbitrary initial state to any arbitrary final state in the absence of noise. The steps described so far are discussed in detail in Sections \ref{sec:equivalence}-\ref{subsec:thm}.

\begin{figure}[hbt!]
    \centering    \includegraphics[width=0.72\linewidth]{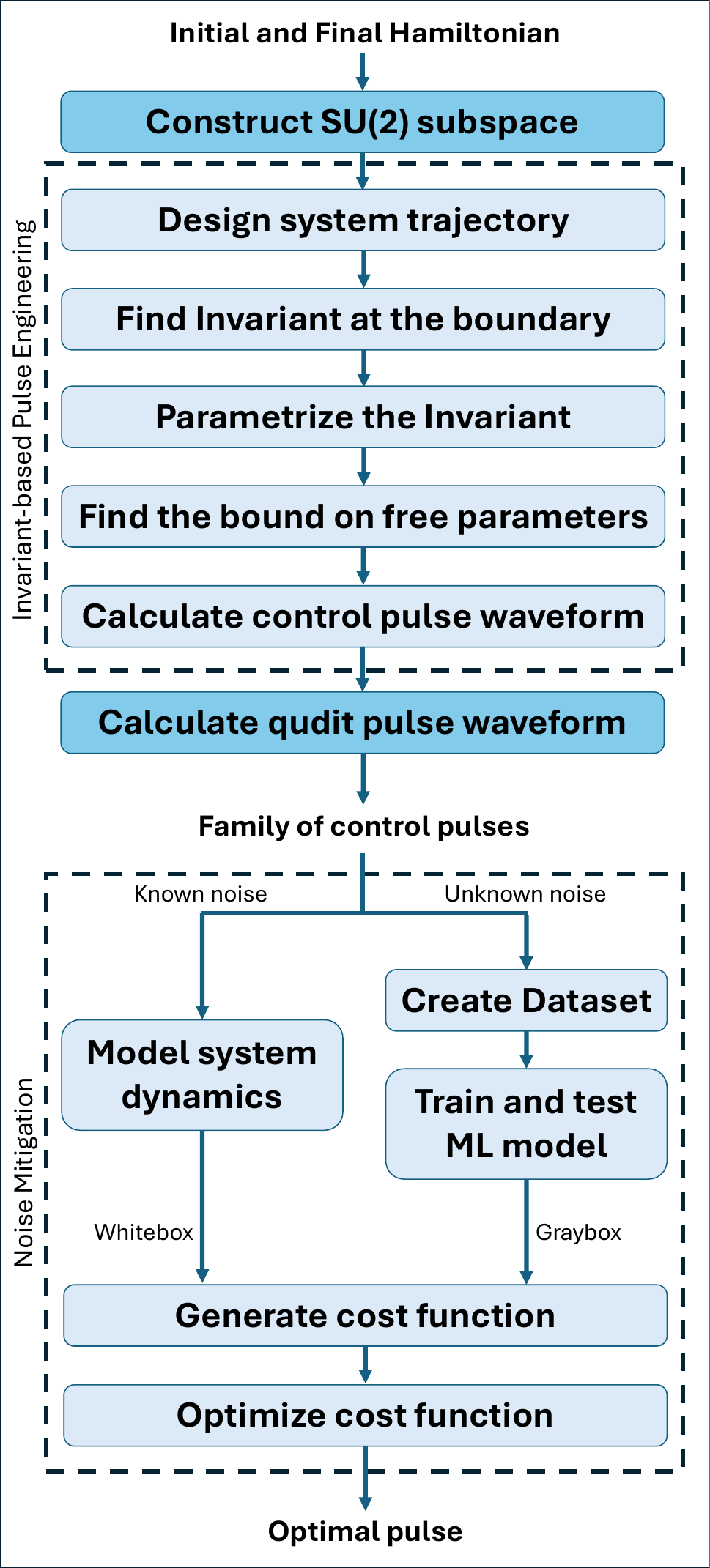}
    \caption{\textbf{The workflow of the proposed method}. The first step in the protocol is to construct an SU(2) subspace, based on the initial and target states effectively reducing the problem to singe-qubit control. Next, we split the target trajectory into sub-trajectories to avoid singularities in control pulses. After that, the invariant is defined at the boundary points for each subtrajectory, followed by finding a parameterized functional form of the invariant, and bounding the parameters to avoid complex-valued pulses. The corresponding family of control pulses of the SU(2) subspace  can then be computed, achieving the target evolution in the absence of noise. Those pulses can then be transformed back to obtain the physical qudit control pulses. The next step, is finding the optimal control pulse from the constructed family to mitigate the noise effects. If the noise model is known, a control cost function is optimized directly. Otherwise, a machine learning stage is introduced where an ML model is designed and trained on a dataset. The trained ML model can then be integrated into the cost function to find the optimal control.
    }
    \label{fig:flowchart}
\end{figure}

Finally, we address the presence of noise affecting the system. If the mathematical description of the noise is known, we directly optimize a cost function, to find the pulse that minimizes the noise effects from the constructed control pulse family. On the other hand, if the noise is unknown, or mathematically intractable to express, then we introduce an additional machine learning stage, to model such noise. This will require constructing an experimentally-accessible dataset to train a machine-learning model, which can then be utilized in the cost function to find the optimal control pulse. Our proposed machine-learning design allows addressing any general open-system dynamics including the non-Markovian case. We discuss noise mitigation in detail in Section \ref{sub2}. Figure \ref{fig:flowchart} summarizes the workflow of our proposed methodology. 

\begin{figure*}[hbt!]
    \centering    \includegraphics[width=0.8\linewidth]{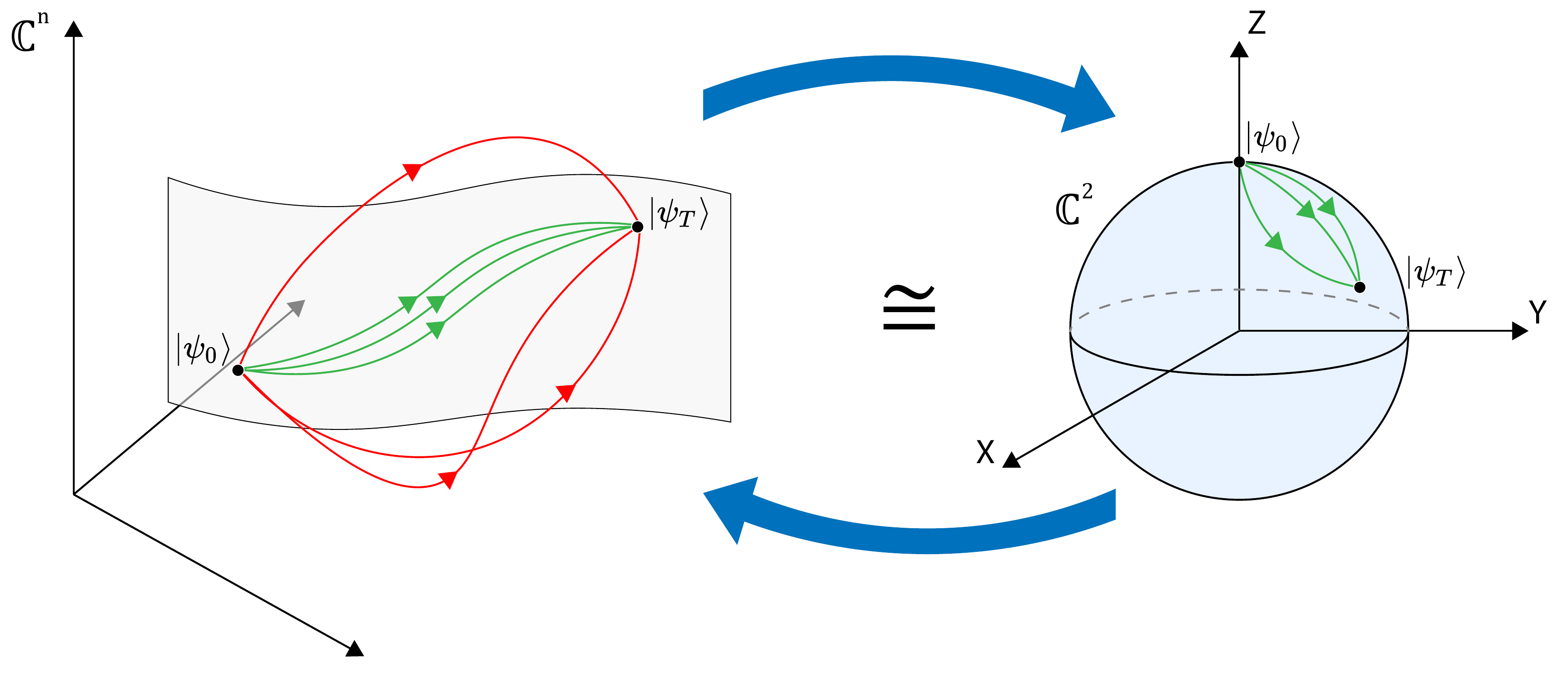}
    \caption{\textbf{Mapping of a finite-dimensional space onto an SU(2) subspace}. We construct an SU(2) subspace that contains the initial and target states. We restrict the closed-system evolution to that subspace, effectively discarding any trajectories that are not fully contained. The qudit state preparation problem can then be reduced to a single-qubit preparation within this subspace. After the SU(2) pulses are designed, the qudit control Hamiltonian can be constructed by the inverse mapping, and can be represented in any relevant set of basis.
    }
    \label{fig:mapping}
\end{figure*}

\subsection{Constructing SU(2) subspace}\label{sec:equivalence}

The control problem of preparing a target state in a finite-dimensional closed system can be transformed into an equivalent problem of state preparation on a single qubit. Let 
$\ket{\psi_{\text{in}}}$ and $\ket{\psi_{\text{tar}}}$ be the initial and target states of a $d$-dimensional qudit. There exist an infinite number of SU(2) subspaces that can contain both the initial and target states. Here, we choose the following SU(2) subspace:
\begin{align}\label{eq:subdef}
    \mathcal{S}=\text{span}\left\{\ket{0_S},\ket{1_S}\right\},
\end{align}
where 
\begin{align}
    \ket{0_S}&:=\ket{\psi_{\text{{in}}}}\\
    \ket{1_S}&:=\ket{\psi_{\text{{in}}}^\perp}\\
    &=\frac{1}{N}\left(\ket{\psi_{\text{tar}}}-\braket{\psi_{\text{in}}|\psi_{\text{tar}}}\ket{\psi_{\text{in}}}\right).
\end{align}
Here, $\ket{\psi_{\text{{in}}}^\perp}$ is the component of the target state that is orthogonal to the initial state, and $N$ is the normalization constant to ensure $\braket{1_S|1_S}=1$. The states $\ket{0_S}$ and $\ket{1_S}$ are clearly orthonormal. We can express the initial and target states as 
\begin{align}
    \ket{\psi_{\text{{in}}}}&=\ket{0_S}\\
    \ket{\psi_{\text{tar}}}&=\braket{\psi_{\text{in}}|\psi_{\text{tar}}}\ket{0_S}+\braket{\psi^\perp_{\text{in}}|\psi_{\text{tar}}}\ket{1_S}.
\end{align}
We can then define a set of operators
\begin{align}
    \Sigma_x&=\ket{1_S}\!\bra{0_S}+\ket{0_S}\!\bra{1_S}\label{sigx}\\
    \Sigma_y&=i\ket{1_S}\!\bra{0_S}-i\ket{0_S}\!\bra{1_S}\label{sigy}\\
    \Sigma_z&=\ket{0_S}\!\bra{0_S}-\ket{1_S}\!\bra{1_S}\label{sigz},
\end{align}
which can be easily shown to follow the standard properties of qubit Pauli operators:
\begin{align}
    \text{Tr}[\Sigma_j]&=0\\
    \Sigma_j^\dagger&=\Sigma_j\\
    [\Sigma_j,\Sigma_k]&=2i\epsilon_{jkl}\Sigma_l\\
    \Sigma_j^2&=\ket{0_S}\!\bra{0_S} + \ket{1_S}\!\bra{1_S}\\
    &:=I_\mathcal{S}\\
    \frac{1}{2}\text{Tr}[\Sigma_j^\dagger\Sigma_k]&=\delta_{jk},
\end{align}
where $j,k,l \in \{x,y,z\}$, $\epsilon_{jkl}$ is the Levi-Civita symbol. In other words, as a consequence of the orthonormality of $\{\ket{0_S}, \ket{1_S}\}$, these $d\times d$ operators are traceless, Hermitian, closed under commutation, square to the subspace identity, and orthonormal under the Hilbert-Schmidt inner product. Because of the assumption of having full qudit control, we can choose the control pulses such that the control Hamiltonian is restricted to the form  
\begin{align}
    H_{\text{ctrl}}(t) = \tfrac{1}{2}(h_x(t)\Sigma_x + h_y(t)\Sigma_y + h_z(t)\Sigma_z).
    \label{h_SIGMA}
\end{align}
We can then show that in the absence of noise and given any state of the form $\alpha_0\ket{0_S} + \alpha_1\ket{1_S}$ lying within $\mathcal{S}$, the evolved state remains in $\mathcal{S}$ under the action of any unitary matrix defined using that Hamiltonian. This allows us then to represent the dynamics as a single-qubit system by identifying the mapping,
\begin{align}
    \ket{0_S}_{d\times 1} &\mapsto \ket{0}_{2\times 1}\\
    \ket{1_S}_{d\times 1} &\mapsto \ket{1}_{2\times 1}\\
    (\Sigma_{x})_{d\times d} &\mapsto (\sigma_x)_{2\times 2}\\
    (\Sigma_y)_{d\times d} &\mapsto (\sigma_y)_{2\times 2}\\
    (\Sigma_z)_{d\times d} &\mapsto (\sigma_z)_{2\times 2},
\end{align}
effectively reducing the problem to single-qubit control. Now, the goal is finding the control pulses $h_j(t)$ such that $2\times1$ initial and final states are given by
\begin{align}
    \ket{\psi_0}&=\ket{0}\\
    \ket{\psi_T}&=\braket{\psi_{\text{in}}|\psi_{\text{tar}}}\ket{0}+\braket{\psi^\perp_{\text{in}}|\psi_{\text{tar}}}\ket{1}.
\end{align}
The initial and final Hamiltonians that have these states as the ground state can be easily obtained as 
\begin{align}
    H(0) &= -\sigma_z\\
    H(T) &= \ket{\varphi_T}\!\bra{\varphi_T}-\ket{\psi_T}\!\bra{\psi_T},
\end{align}
where $\ket{\varphi_T} = \braket{\psi^\perp_{\text{in}}|\psi_{\text{tar}}}\ket{0}-\braket{\psi_{\text{in}}|\psi_{\text{tar}}}\ket{1}$. Now, the problem is fully mapped to single-qubit control, and in the following section we focus on how to design the pulses ensuring boundedness at all points of time. Figure \ref{fig:mapping} illustrates the aforementioned proposal.

\subsection{Designing system trajectory}\label{subsec:traj}

Previous works that employ dynamical invariants \cite{levy2018noise,invopen1,invcat} usually utilize a functional form of the invariants that leads to  control pulses expressed as rational trigonometric functions. This can lead to unbounded pulses in some cases (i.e. pairs of initial/target states), limiting its applicability in a real experimental setting. This happens due to the relation between the invariant coefficients and the control pulses (for example Equation \ref{eq:drift}), which always involves a rational expression. Because of the boundary conditions that have to be imposed on the invariant, there will be cases where the denominator of such expression vanishes, leading to a singularity at some point. In order to address this issue, we split the trajectory between the initial state and the desired target state into subtrajectories such that there the denominator does not vanish at any point during evolution. In this section we will show how to do this splitting.

Since the relations in Equation \ref{fhrelation} are linearly-dependent, there exist an infinite number of Hamiltonians corresponding to a given invariant. Therefore, we can pick one Hamiltonian by fixing one of its coefficients $h_x(t)$, $h_y(t)$, or $h_z(t)$, which we refer to as the \textit{reference pulse}, and denote it by $h_3(t)$, while referring to the corresponding axis $\sigma_3$ as the \textit{reference axis}. Based on the selection, we introduce the alternative notation for the control Hamiltonian and Invariant,
\begin{align}\label{eq:notation}
    I(t) &= \frac{f_1(t)}{2}\sigma_1 + \frac{f_2(t)}{2}\sigma_2 + \frac{f_3(t)}{2}\sigma_3\\
    H_{\text{ctrl}}(t) &= \frac{h_1(t)}{2}\sigma_1 + \frac{h_2(t)}{2}\sigma_2 + \frac{h_3(t)}{2}\sigma_3. 
\end{align}
The pulses $h_1(t) $ and $h_2(t)$ will then refer to one of the two remaining axes according to Table \ref{table:pulse_calc}. For example, if the reference axis is $Y$, then $h_3(t)=h_y(t)$, while $h_1(t)=h_z(t)$ and $h_2(t)=h_x(t)$. This notation applies identically to the corresponding invariant coefficients $f_x(t)$, $f_y(t)$, and $f_z(t)$. The functional form of the reference pulse can be chosen arbitrarily, from which the other two pulses can be computed as, 
\begin{align}\label{eq:pulse_calc}
    h_1(t) &= \frac{f_1(t)h_3(t) - \dot{f}_2(t)}{f_3(t)}\nonumber\\
    h_2(t) &= \frac{f_2(t)h_3(t) + \dot{f}_1(t)}{f_3(t)}.
\end{align}
The key to avoiding the vanishing of the denominator in Equation \ref{eq:pulse_calc} is to choose a different reference axis for each subtrajectory. The number of subtrajectories and their corresponding reference axis will depend on initial and target Hamiltonian as follows. For the purpose of this paper, we assume that the initial state is $\ket{\psi(0)}=\ket{0}$ or $H_{\text{ctrl}}(0)=-\sigma_z$, which is easy to prepare in most quantum systems. This sets $f_x(0)=f_y(0)=0$ so the reference axis has to be $Z$, for the first subtrajectory. Now depending on the target state location on the Bloch sphere, four different scenarios can arise:

\begin{itemize}[leftmargin=*]
\item{\textbf{Case 1:}} The target state is in the same hemisphere as the initial state, or $h_z(0)h_z(T)>0$. We select the reference axis to be $Z$ at all times (i.e. no splitting required). Moreover, we can choose the final Hamiltonian such that $h_z(t)=h_z(0)$ (i.e. a constant function), during the evolution. This can also be applied in systems with constant drift and control present only along $X$ and $Y$ direction.
    
\item{\textbf{Case 2:}} The target state is either on the equator, or on the other hemisphere, i.e. $h_z(0)h_z(T)\leq 0$. Additionally, if $h_x(T)\neq 0$, then we have the following two subtrajectories between $[0,T/2]$ and $[T/2,T]$ with the intermediate Hamiltonian chosen to be $H_{\text{ctrl}}(T/2)=h_x(T)\sigma_x+\frac{1}{2}\left(h_y(0)+h_y(T)\right)\sigma_y + h_z(0)\sigma_z$: 
\begin{enumerate}
        \item Initial state to intermediate state with $Z$ as reference axis and $h_z(t)=h_z(0), \forall t \in [0,T/2]$.
        \item Intermediate state to target state with $X$  as reference axis and $h_x(t)=h_x(T), \forall t\in [T/2,T]$.
\end{enumerate}

\item{\textbf{Case 3:}} Similar to Case 2 where $h_z(0)h_z(T)\leq 0$, however, we have $h_x(T) = 0$ and $h_y(T)\neq 0$. Here we define two subtrajectories between $[0,T/2]$ and $[T/2,T]$ with the intermediate Hamiltonian chosen to be $H_{\text{ctrl}}(T/2)=\frac{1}{2}\left(h_x(0)+h_x(T)\right)\sigma_x+h_y(T)\sigma_y + h_z(0)\sigma_z$:
\begin{enumerate}
        \item Initial state to intermediate state with $Z$ as reference axis, with $h_z(t)=h_z(0), \forall t \in [0,T/2]$.
        \item Intermediate state to target state with $Y$ as reference axis, with $h_y(t)=h_y(T), \forall t\in [T/2,T]$.
\end{enumerate}
    
\item \textbf{Case 4:} The target state is antipodal with respect to initial state, or $h_z(0)h_z(T)\leq 0$, $h_x(T) = 0$ and $h_y(T) = 0$. Here we need to construct three subtrajectories between $[0,T/3]$, $[T/3,2T/3]$, and $[2T/3,T]$. The intermediate Hamiltonians are chosen to be $H_{\text{ctrl}}(T/3)=h_z(0)\sigma_x+h_y(T)\sigma_y + h_z(0)\sigma_z$ and $H_{\text{ctrl}}(2T/3)=h_z(0)\sigma_x+h_y(T)\sigma_y + h_z(T)\sigma_z$:
    \begin{enumerate}
        \item Initial state to first intermediate state with $Z$ as reference axis and $h_z(t)=h_z(0), \forall t \in [0,T/3]$.
        \item First Intermediate state to second intermediate state with $X$ as reference axis and $h_x(t)=h_z(0), \forall t\in [T/3,2T/3]$.
        \item Second intermediate state to target state with $Z$ as reference axis and $h_z(t)=h_z(T), \forall t \in [2T/3,T]$.
    \end{enumerate}
\end{itemize}

\begin{figure*}
    \centering
    \includegraphics[width=\linewidth]{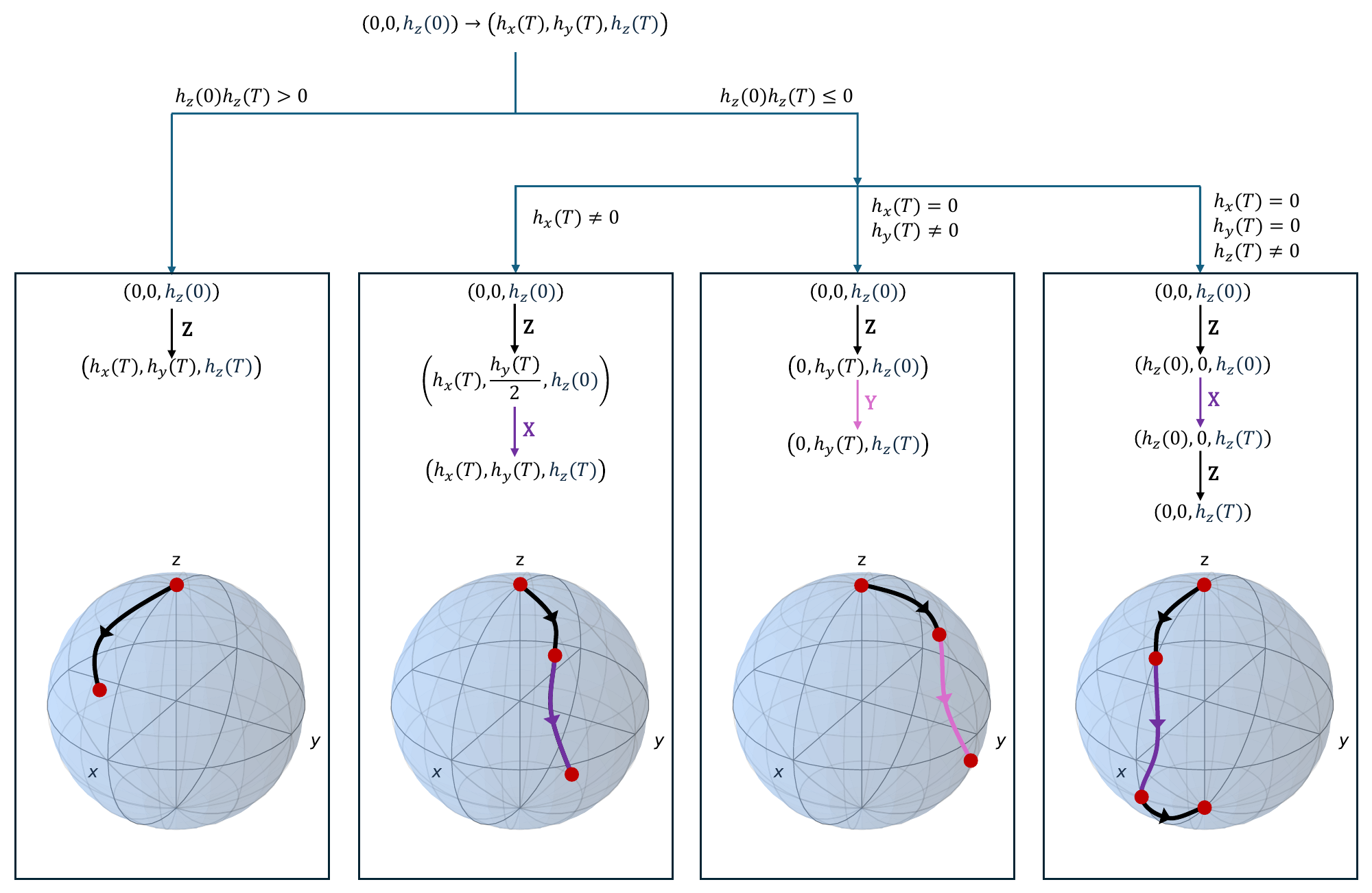}
    \caption{\textbf{Trajectory splitting to avoid singular (unbounded) pulses.} Given the initial and target control Hamiltonians,  $H_{\text{ctrl}}(0)=h_z(0)\sigma_z/2$ and $H_{\text{ctrl}}(T)=(h_x(T)\sigma_x+h_y(T)\sigma_y + h_z(T)\sigma_z)/2$, respectively, we split the trajectory into a number of subtrajectories, based on the location of the target state. This step of the protocol is designed to avoid the control pulse amplitude from growing to infinity. For each subtrajectory, a reference axis is chosen alongside the corresponding intermediate points.  
    }
    \label{fig:design_traj}
\end{figure*}

These four cases cover all possible target states for a single qubit, starting from the initial state $\ket{0}$. Figure \ref{fig:design_traj} summarizes these four cases. Similar logic can be followed to design the subtrajectories given a different initial state. The outcome of this step is finding the appropriate boundary Hamiltonian for each subtrajectory. 

\begin{table}[t]
\caption{\textbf{Axis labeling in relation to the choice of reference axis.} Depending on the choice of a reference axis, the labeling of the Pauli matrices, the coefficients of the control Hamiltonian, and the coefficient of the Invariant are determined correspondingly.}
\label{table:pulse_calc}
\centering
\begin{tabular}{|c|c|c|c|}
\hline
\textbf{Ref. axis} & $\mathbf{\sigma_1}$ & $\mathbf{\sigma_2}$ & $\mathbf{\sigma_3}$ \\
\hline
X & $\sigma_y$ & $\sigma_z$ & $\sigma_x$ \\
Y & $\sigma_z$ & $\sigma_x$ & $\sigma_y$\\
Z & $\sigma_x$ & $\sigma_y$ & $\sigma_z$\\ 
\hline
\end{tabular}
\end{table}

\subsection{Defining the boundary invariants}\label{sec:boundary}
The next step is to find the boundary invariant corresponding to the boundary Hamiltonians of each subtrajectory. Given a subtrajectory that is defined over $ t \in [t_i,t_f]$, we require that $[I(t_b),H_{\text{ctrl}}(t_b)] = 0$, where $t_b \in \{t_i,t_f\}$. This condition ensures that the system evolves from the ground state of the $H_{\text{ctrl}}(t_i)$ to the ground state of $H_{\text{ctrl}}(t_f)$. This is necessary for the endpoints $t_i=0$ and $t_f=T$, to guarantee that the system evolves to the target state at $t=T$. On the other hand, this condition can be relaxed at the intermediate points in general. However, imposing it provides an advantage in terms of the final pulse design, which is ensuring the continuity of the pulse across subtrajectories. In particular, from Equation \ref{eq:bc1}, the commutation with the Hamiltonian implies that $\dot{f}_k(t_b)=0, \ k\in\{1,2,3\}$, which holds for the final point of a given subtrajectory and the initial point of the next one. In other words,
\begin{align}
    \lim_{t\to t_b^+}\dot{f}_k(t) = \lim_{t\to t_b^-}\dot{f}_k(t) = 0.
\end{align}
Thus, if the functional form of the invariant is chosen to be a differentiable function in each subtrajectory, then the invariant is differentiable at every point. Referring to Equation \ref{eq:pulse_calc}, since $f_3(t)$ is non-vanishing and continuous, then all the terms in the expression are continuous. We can then conclude that $h_1(t)$ and $h_2(t)$ will be continuous over $[0,T]$. In an experimental setting, continuous control pulses are generally more favorable compared to discontinuous pulses as they occupy lower bandwidth.

The invariant and the control Hamiltonian, as represented in Equation \ref{eq:notation}, would commute only if 
\begin{equation}\label{eq:ratio}
    \frac{f_1(t_b)}{h_1(t_b)} = \frac{f_2(t_b)}{h_2(t_b)} = \frac{f_3(t_b)}{h_3(t_b)} = k(t_b),
\end{equation}
where $k(t_b)$ is a constant for each boundary of each subtrajectory. Substituting this in Equation \ref{eq:sphere}, we get
\begin{equation}
    k(t_b)^2\left(h_1(t_b)^2 + h_2(t_b)^2 + h_3(t_b)^2\right) = c^2
\end{equation}
or 
\begin{equation}
    k(t_b) = \frac{c}{\sqrt{h_1(t_b)^2 + h_2(t_b)^2 + h_3(t_b)^2}}.
\end{equation}
So this fixes the invariant at the boundaries of each subtrajectory in terms of control Hamiltonian. To sum up, we have 4 boundary conditions for the invariant coefficients:
\begin{align}\label{eq:boundary}
    f_1(t_b) &= \frac{ch_1(t_b)}{\sqrt{h_1(t_b)^2 + h_2(t_b)^2 + h_3(t_b)^2}}\nonumber\\
    f_2(t_b) &= \frac{ch_2(t_b)}{\sqrt{h_1(t_b)^2 + h_2(t_b)^2 + h_3(t_b)^2}}\\
    \dot{f}_1(t_b) &= \dot{f}_2(t_b) = 0.\nonumber
\end{align}
The final invariant coefficient $f_3(t)$ is fixed due to Equation \ref{eq:sphere}, and thus there is no need to specify explicitly its boundary conditions.

\subsection{Parameterizing the invariant}\label{subsec:invdgn}
The third step in our protocol is to find a suitable functional form for the invariant coefficients for each subtrajectory, which satisfy the boundary conditions given earlier. In this paper, we choose $f_1(t)$ and $f_2(t)$ to be $n^{\text{th}}$-degree polynomials of the form
\begin{equation}
    f_k(t) = a_{k,0} + a_{k,1}t + a_{k,2}t^2 + \dots + a_{k,n}t^{n}
\end{equation}
for $k=1,2$. $f_3(t)$ is then computed using Equation \ref{eq:sphere}
\begin{equation}\label{eq:f3}
    f_3(t) = \pm\sqrt{c^2 - f_1^2(t) - f_2^2(t)}.
\end{equation}
Here, we select the sign of $f_3(t)$ to be the same as the sign of $h_3(t)$ in each subtrajectory. Note that, the reference pulse $h_3(t)$ has been chosen to constant in each subtrajectory as previously discussed in Section \ref{subsec:traj}. This also results in the invariant and Hamiltonian having the same ground state at the boundaries. 

To simplify the design process and improve numerical precision of the computations, we work in normalized time $s = (t-t_i)/(t_f-t_i)$, so that the interval $t\in[t_i,t_f]$ gets mapped to the interval $s \in [0,1]$. In this normalized time range, the polynomials take the form $\tilde{f}_k(s) = a_{0,k} + a_{1,k}s + a_{2,k}s^2 + \dots + a_{n,k}s^{n}$. These can be expressed in vector form as 
\begin{align}
    \tilde{f}_k(s) &= \begin{pmatrix}
        1 & s & \cdots & s^{n}
    \end{pmatrix}\begin{pmatrix}
        a_{k,0} & a_{k,1} & \cdots & a_{k,n}
    \end{pmatrix}^T\\
    &:=\xi^T(s)\vec{x}_k
\end{align}
If we express the boundary conditions for these polynomials in matrix form, we obtain a system of linear equations of the form $A\vec{x}_k = \vec{b}_k$, where 
\begin{align}
    A &= \begin{pmatrix}
        \xi^T(0) \\ \dot{\xi}^T(0) \\ \xi^T(1) \\ \dot{\xi}^T(1)
    \end{pmatrix}=\begin{pmatrix}
    1 & 0 & 0 & \cdots & 0\\
    0 & 1 & 0 & \cdots & 0\\
    1 & 1 & 1 & \cdots & 1\\
    0 & 1 & 2 & \cdots & n\\
\end{pmatrix}_{4\times (n+1)},
\end{align} 
and $\vec{b}_k = \begin{pmatrix}  \tilde{f}_k(0) & 0 &     \tilde{f}_k(1) & 0  \end{pmatrix}^T$. In order to satisfy these four conditions, in general these polynomials should be at least of degree 3. In this case, the solution is given by 
\begin{align}
    \vec{x}_k^{(3)} = \begin{pmatrix}
        \tilde{f}_k(0)\\
        0\\
        3(\tilde{f}_k(1) - \tilde{f}_k(0))\\
        -2(\tilde{f}_k(1) - \tilde{f}_k(0))
    \end{pmatrix}.
\end{align}
Using Equations \ref{eq:boundary}, we can obtain the function form for the invariant coefficient in normalized time,
\begin{align}
    \tilde{f}_{k}(s) =& \frac{ch_k(t_i)}{R(t_i)} + c\left(\frac{h_k(t_f)}{R(t_f)} - \frac{h_k(t_i)}{R(t_i)} \right)(3s^2-2s^3),
\end{align}
where $R(t) = \sqrt{h_1^2(t) + h_2^2(t) + h_3^2(t)}$. However, if a polynomial of higher degree is chosen, then an infinite number of solutions for $\vec{x}_k^{(n)}$ will exist satisfying the system of linear equations. This creates a parameterized family of solutions converging at the boundary, $s\in\{0,1\}$, while differing at intermediate points $s\in(0,1)$. This provides flexible intermediate dynamics corresponding to the invariant, while maintaining the necessary boundary conditions.  Moreover, higher-degree polynomials will exhibit more shape variations, and thus can provide more flexibility in system dynamics. This is crucial when dealing with noise mitigation in subproblem 2 as will be discussed later. In the general case where $n>3$, the solution of the linear system of equations is of the form
\begin{align}\label{eq:polyvec}
    \vec{x}_k^{(n)} &= c\vec{e}_k + \sum_{j=1}^{m} v_{k,j} \vec{u}_j\\
     &= \begin{pmatrix}
        \vec{e}_k & \vec{u}_1 & \cdots & \vec{u}_m
    \end{pmatrix}\begin{pmatrix}
        c & v_{k,1}  & \cdots & v_{k,\text{m}}\end{pmatrix}^T\\
    &:= M_k\vec{\theta}_k,
\end{align}
where 
\begin{equation} \label{eq:ek}
    c\vec{e}_k = \begin{pmatrix}\vec{x}_k^{(3)}  \\ \vec{\mathbf{0}}_{m} \end{pmatrix}
\end{equation}
is the concatenation of the third degree solution and the zero vector of dimensions $m$, $\vec{u}_i$ is the
$i^{\text{th}}$ null vector of $A$, $m=\text{nullity}(A)=n-3$, and $v_{k,i}$ is an arbitrary real number. The polynomial matrix $M_k$ is of size ${n\times (\text{m}+1)}$. It can be obtained numerically by computing $\vec{e}_k$ as in Equation \ref{eq:ek}, and concatenating it with the null vectors $\vec{u}_j$. The null vectors can be calculated by performing singular value decomposition (SVD) on $A$ for any degree $n$.

The vector of coefficients $\vec{\theta}_k$ consists of the fixed element $c$ in its first entry, followed by the elements $v_{k,i}$ which are treated as free variables. The degree 3 solution can then be recovered if we set $v_{k,i}=0, \ \forall k,i$. The free variable $v_{k,i}$ hence define the parameterization of the family of invariants.
The polynomials can then be expressed as
\begin{equation}\label{matrixtopoly}
    \tilde{f}_k(s) = \xi^T(s) M_k  \vec{\theta}_k.
\end{equation}
This form enable efficient representation of the polynomials which facilitate performing optimization in Subproblem 2. In that case, the optimization algorithm is designed to find a specific value of $\vec{\theta}_k$ which minimizes a given cost function. 

\subsection{Bounding the free variables}\label{subsec:vmax}
As discussed previously, the polynomial of the form \ref{matrixtopoly} would satisfy the boundary conditions with any value of the free variables $v_{k,i}$. However, these free variables need to be constrained to a maximum value $v_{\max}$ to avoid having a complex-valued $f_3(t)$ as per Equation \ref{eq:f3}. Otherwise the control pulses will become complex-valued which is non-physical. So, the next step in the workflow is finding the maximum coefficient of the null vectors ($v_{\max}$), i.e. $|v_{k,j}|\le v_{\max}$ such that
\begin{align}\label{eq:ineq}
     f_3^2(t)=c^2-f_1^2(t)-f_2^2(t)>0, \ \forall t \in [t_i,t_f].
\end{align}
Finding $v_{\max}$ analytically to satisfy this inequality might be challenging. Here we propose a simpler heuristic approach to find the solution. The idea is based on finding the maximum allowed value of the free variable locally $v_{\max}(s)$ for each $s\in[0,1]$, then find the minimum over the whole normalized time range. In other words,
\begin{align}
    v_{\max} = \min_{s\in[0,1]}v_{\max}(s).
\end{align}
Now to find $v_{\max}(s^*)$ for a given $s^*\in[0,1]$, we start by rewriting the invariant coefficients in Equation \ref{eq:polyvec} in polynomial form parameterized by the free variables $v_{k,j}$,
\begin{equation}\label{polyexpansion}
    \tilde{f}_k(s^*|v_{k,1},\cdots, v_{k,m}) = c{e}_k(s^*) + \sum_{j=1}^{m} v_{k,j}{u}_j(s^*),
\end{equation}
where ${e}_k(s^*)=\xi^T(s^*)\vec{e}_k$ and ${u}_j(s^*)=\xi^T(s^*)\vec{u}_j$. This expression is in the form of a liner combination of the free variables $v_{k,j}$, which are all bounded $-v_{\max}(s^*)\le v_{k,j} \le v_{\max}(s^*)$. The maximum absolute value of this combination is obtained when all the free variables are $\pm v_{\max}(s^*)$, with their signs chosen such that all the terms in the combination have the same sign, or
\begin{equation}
    v^{*}_{k,j} = \begin{cases}
        v_{\max}(s^*) & u_j(s^*)e_k(s^*) \geq 0\\
        -v_{\max}(s^*) & u_j(s^*)e_k(s^*) < 0.
    \end{cases}
\end{equation}
If $e_k(s^*)=0$ at some point, then the signs of $v_{k,j}$ are chosen in this case to be the same as $u_j(s^*)$, or  
\begin{equation}
    v^{*}_{k,j} = \begin{cases}
        v_{\max}(s^*) & u_j(s^*) \geq 0\\
        -v_{\max}(s^*) & u_j(s^*) < 0.
    \end{cases}
\end{equation}
We substitute these coefficients back in \ref{polyexpansion}, and represent it in the form
\begin{align}
\label{eq:fen}
    \tilde{f}_k(s^*|v^*_{k,1},\cdots, v^*_{k,m}) &= c{e}_k(s^*) + v_{\max}(s^*)n_{k}(s^*) \\
    &:=g_k(v_{\max}(s^*)),
\end{align}
where $n_k(s^*) = \sum_{j=1}^{m} \sgn(v_{k,j}^*)u_j(s^*)$. This gives a tight bound on the invariant coefficient at $s^*$,
\begin{align}\label{eq:fmax}
    | \tilde{f}_k(s^*)| \le  |g_k(v_{\max}(s^*))|.
\end{align}
Going back to Equation \ref{eq:ineq}, we have
\begin{align}
   \tilde{f}_3^2(s^*) &= c^2 - \tilde{f}_1^2(s^*)-\tilde{f}_2^2(s^*) \\
   &=c^2 - |\tilde{f}_1(s^*)|^2-|\tilde{f}_2(s^*)|^2 \\
   &\ge c^2 -  |g_1(v_{\max}(s^*))|^2 - |g_2(v_{\max}(s^*))|^2 
\end{align}
The third step is from substitution of Equation \ref{eq:fmax}. This gives the required condition on $v_{\max}$, which is ensuring the RHS of the last line to be strictly positive, or, 
\begin{equation}\label{cexpr}
    c^2 - |g_1(v_{\max}(s^*))|^2 - |g_2(v_{\max}(s^*))|^2 > 0.
\end{equation}
Using Equation \ref{eq:fen}, and rearranging the terms, we get that the required condition is
\begin{equation}\label{eq:quad}
    - \left(\frac{v_{\max}(s^*)}{c}\right)^2C_1 - \frac{2v_{\max}(s^*)}{c}C_2 + C_3> 0
\end{equation}
where $C_1=n_{1}^2(s^*) + n_{2}^2(s^*)$, $C_2(s^*)=e_{1}(s^*)n_{1}(s^*) + e_{2}(s^*)n_{2}(s^*)$ and $C_3=1 - e_{1}^2(s^*) - e_{2}^2(s^*)$. In Supplementary Note 3, we show that $C_3>0, \forall s\in[0,1]$. Notice that the LHS of the required inequality is a quadratic function of $v_{\max}(s^*)/c$, with the coefficient of the quadratic term $-C_1$ being strictly negative. Thus, it is a concave down function. Moreover, we observe that setting $v_{\max}(s^*)=0$, satisfies the inequality, showing the existence of at least one positive value of the LHS. These observations lead to the conclusion that the quadratic must have two real roots. Since, the product of these two roots is $C_3/-C_1$, which is strictly negative, this implies that the two solutions must have opposite signs. Therefore, the inequality is satisfied for any $v_{\max}(s^*)/c \in (V_1(s^*), V_2(s^*))$, where $V_1(s^*)<0$ and $V_2(s^*)>0$ are the roots. Since $v_{max}(s^*)$ is defined to be a bound on the absolute value of free variables, then a negative solution is not allowed. So we choose the greatest upper bound possible which is $v_{\max}(s^*)<V_2(s^*)$. 

Now that we have found an upper bound for $v_{\max}(s^*)$ at each point $s^*\in [0,1]$, we find the minimum over all times, to obtain the global $v_{\max}$. This min-max approach would guarantee that the inequality holds at each point of the dynamics, implying that $\tilde{f}_3(s)>0,\ \forall s \in [0,1]$. This procedure can be performed numerically by discretizing the normalized time interval $s \in [0,1]$, and finding the positive root $V_2(s^*)$ of the LHS in \ref{eq:quad} at each point $s^*$. Then, we take the minimum over all points, and rescale by a factor of $1-\epsilon$ for a small positive number $\epsilon\to 0$. In other words, we set $v_{\max}=(1-\epsilon)\min_{s^*}V_2(s^*)$. This ensures that numerically the inequality in Equation \ref{eq:quad} holds strictly even for $v_{k,j}=\pm v_{\max}$. 

The higher the magnitudes of the free variables, the more variations will be present in the family of control pulses. This can be advantageous when addressing subproblem 2. On the other hand, values of the free variables that are close to $v_{\max}$ will result in control pulses of high amplitudes at some points in time. This is because in Equation \ref{eq:pulse_calc}, the denominator has $f_3(t)$ which will be close to vanishing in that case. Therefore, in practical application of the method, we might need to restrict the domain of the free variables to a smaller range, to limit the amplitudes of the resulting control pulses.

\subsection{Proving the physicality of the control pulses}\label{subsec:thm}
\begin{theorem} 
Splitting the trajectory as introduced in Section \ref{subsec:traj}, ensuring that $h_3(t_i)h_3(t_f)>0$ for each subtrajectory defined over $t\in[t_i,t_f]$, and utilizing the parametrization introduced in Section \ref{subsec:invdgn} with bounds found in Section \ref{subsec:vmax} ensures that $h_1(t),h_2(t)<\infty,\ \forall t \in [0,T]$.
\end{theorem}

\begin{proof}
Let $h_3(t_i)h_3(t_f)>0$. According to Equation \ref{eq:ratio}, we get
\begin{equation}
f_3(t_i)f_3(t_f)=k(t_i)k(t_f)h_3(t_i)h_3(t_f)>0
\end{equation}
as $k(t_i),k(t_f)>0$. This implies that $f_3(t)$ is of the same sign at the boundaries of the subtrajectories. While at intermediate points, its value is computed using Equation \ref{eq:f3}. Bounding the free variable according to section \ref{subsec:vmax} ensure that $f_3(t)$ is always real and non-vanishing. This requires that $h_3(t_i),h_3(t_f)\neq0$, which is automatically satisfied by the assumption. Now choosing the sign in Equation \ref{eq:f3} such that it remains the same for $t\in [t_i,t_f]$ ensures that $f_3(t)$ remains continuous $\forall[t_i,t_f]$. Computing $h_1(t)$ and $h_2(t)$ according to Equation \ref{eq:pulse_calc} gives $h_1(t),h_2(t)<\infty,\ \forall t \in [t_i,t_f]$. Applying the same logic for each subtrajectory, we ensure that $f_3(t)\neq 0\ \forall t\in [0,T]$ and therefore $h_1(t),h_2(t)<\infty,\ \forall t \in [0,T]$.
\end{proof}

On the other hand, if we restrict the evolution to follow one trajectory for all initial/target states (as presented in previous works), or a subtrajectory is designed such that $h_3(t_i)h_3(t_f)<0$, then we cannot avoid singular (unbounded) pulses. In these situations, we have $f_3(t_i)f_3(t_f)<0$ following Equation \ref{eq:ratio}, which means $f_3(t)$ changes sign in between. Note that $f_3(t)$ is the square-root of a polynomial, following Equation \ref{eq:f3}, and thus is continuous. Following the intermediate value theorem, 
\begin{equation}
    \exists t^*\in [t_i,t_f]: f_3(t^*)=0,
\end{equation}
resulting in $h_1(t^*), h_2(t^*)\to\infty$, i.e, singular control pulses. For example, for the case of population inversion \cite{levy2018noise}, we have $h_x(0)=h_y(0)=h_x(T)=h_y(T)=0$. If we set $X$ or $Y$ as the reference axis, we will get $f_3(0)=f_3(T)=0$, resulting in unbounded pulses at the beginning and end of evolution time. Alternatively, if we consider $Z$ as the reference axis, we have $h_z(0)h_z(T)<0$, which would result in $h_x(t)$ and $h_y(t)$ going to infinity at least at one point in the evolution time, as proven earlier. This shows the importance of following the strategy presented in Sections \ref{subsec:traj}-\ref{subsec:vmax} for splitting the trajectory and bounding the free variables to ensure obtaining  physical control pulses.

\subsection{Mitigating the noise effects}\label{sub2}
After defining a family of invariants using the protocol discussed earlier, we can obtain a corresponding family of control pulses by fixing the reference coefficient ($h_3$) and applying Equation \ref{eq:pulse_calc} for each subtrajectory. The pulses in each of the $X$, $Y$, and $Z$ axes are then concatenated over all subtrajectories, to obtain one full sequence for each direction. Since the invariant coefficients $f_k(t)$in each subtrajectory depend on the free parameters $\vec{\theta_k}$ according to Equation \ref{matrixtopoly}, the control pulses $h_k(t)$ will also depend on these parameters. We denote the concatenation of the free parameters for all subtrajecotries and directions by $\vec{\Theta}$, which uniquely defines each member of the control pulse family. Next, we obtain the physical qudit pulses $c_j(t)$ in Equation \ref{quditctrl} from the SU(2) subspace pulses $h_{x,y,z}(t)$ by a simple transformation of basis using Equations \ref{sigx} to \ref{sigz}. We can then select one member of the family of control pulses to minimize the noise effects as discussed previously.

In order to obtain such optimal pulse, we optimize a cost function that encodes the performance of a given control pulse. In this paper, we choose the cost function to be the infidelity between the final state in the presence of noise, and the desired target state. This can be defined as
\begin{align}\label{eq:cost}
    J(T) &= 1 - \mathcal{F}(T)\nonumber\\
    &:=1 - \bra{\psi_T}\rho(T)\ket{\psi_T} \nonumber \\
    &= 1 - \bra{\psi_T} \left( \sum_{j=0}^{d^2-1} \Tr[\rho(T) \lambda_j] \lambda_j \right) \ket{\psi_T}\\
    &= 1 - \sum_{j=0}^{d^2-1}\braket{\lambda_j(T)}\braket{\psi_T|\lambda_j|\psi_T},
\end{align}
Here, $\rho(T)$ is final state of the system in the presence of noise, which can be expressed in the form of expectation values of the operators $\{\lambda_j\}$ which we have defined earlier to be Hermitian and orthonormal. This cost function can be easily computed in experimental setups as these expectation values are easily accessible. There are different ways to solve this optimal control problem. In this paper, we apply model-based control, which requires constructing a model $\mathcal{M}$ that maps the control parameters $\vec{\Theta}$ into the expectation values
\begin{equation}
    \braket{\lambda_j(T)} = \mathcal{M}\left(\vec{\Theta}\right).
\end{equation}
If the noise model is known, we can directly optimize the cost function. The expectation values of the Pauli operators can be expressed using the noise operator formalism as in Equation \ref{uvo}. Since the control Hamiltonian is constructed using dynamical invariants, the target state is perfectly achieved in the absence of noise. Thus we have $U_{\text{ctrl}}(T;\vec{\Theta})\rho_S(0) U_{\text{ctrl}}^\dagger(T;\vec{\Theta}) = \ket{\psi_T}\!\!\bra{\psi_T}:=\rho_T$ independent of $\vec{\Theta}$. This reduces the expression of the expectations to 
\begin{equation}\label{eq:voinv}
    \braket{\lambda_j(T)} = \text{Tr}[V_{\lambda_j}(T;\vec{\Theta})\rho_T\lambda_j].
\end{equation}
The noise operator $V_{\lambda_j}(T;\vec{\Theta})$ is dependent on the control Hamiltonian as per Equation \ref{eq:vo} and thus is dependent on $\vec{\Theta}$. The cost function in Equation \ref{eq:cost} can then be expressed as
\begin{equation}\label{eq:Vo_cost}
    J(T;\vec{\Theta}) = \left(1-\sum_{j=0}^{d^2-1}\Tr[V_{\lambda_j}(T;\vec{\Theta})\rho_T\lambda_j]w_j\right),
\end{equation}
where $w_j = \bra{\psi_T}\lambda_j\ket{\psi_T}$, which is also independent of $\vec{\Theta}$. We can then minimize the cost function with respect to $\vec{\Theta}$, to obtain the optimal control pulse. This ``whitebox'' approach is often applied in special cases such as weak coupling, where approximations via perturbative methods can be applied to obtain a tractable expression of $V_{\lambda_j}(T;\vec{\Theta})$ \cite{vo1}.

On the other hand, if the noise model is not completely known, the approximation is not valid, or the expressions are intractable, the whitebox approach would not be suitable. In this case, we propose a more general machine learning approach \cite{W4,W2,qcml1,qcml2}. We construct an ML model with pulse parameters $\vec{\Theta}$ as input and expectations $\braket{\lambda_j(T)}$ as output. The model is trained using a dataset, consisting of pairs of the pulse parameters, and the corresponding expectations of the basis set. This dataset can be easily constructed experimentally, and does not require any inaccessible quantities or information about the unknown noise. The training of the model involves minimizing a loss function (such as Mean Square Error (MSE) between the model prediction and the actual output. Creating a large enough dataset as well as choosing a suitable model architecture is important to ensure high training accuracy while avoiding overfitting. This is essential for the success of the model-based ML approach.

Here, we choose a graybox architecture \cite{W1,W2,W3,W4,W5,W6,gbyule} for the ML model, but other architectures including standard blackbox \cite{wise,qcml1,qcml2} could have been applied. The graybox architecture includes a standard blackbox structures (such as neural networks) as well as physics-based whitebox layers. In our setting, the graybox approximates the relation between the pulse parameters $\vec{\Theta}$ and the expectations as in Equation \ref{eq:voinv}, by designating the blackbox part to approximate the matrix elements of the noise operator $V_{\lambda_j}(T;\vec{\Theta})$. The 
whitebox layers then compute the expression inside the trace in Equation \ref{eq:voinv} to generate the output. A detailed architecture of graybox suitable for qudit systems has been studied in detail in our previous work \cite{gbyule}. The machine learning approach works for any noise environment, and thus avoids the challenges of the purely-whitebox approach. The limitation, however,  is the requirement of creating large-enough datasets and the  computational resources needed. Once the model is trained, it can be used to predict the expectation values $\braket{\lambda_j(T)}$ for any $\vec{\Theta}$ beyond the examples used in the training procedure. This enables computing the cost function in Equation \ref{eq:cost} numerically, and thus optimizing it with respect to the pulse parameters $\vec{\Theta}$. 

The optimization of the cost function is performed using random search algorithm, as explained in Supplementary Note 4. Once the optimal parameters are obtained, the corresponding control pulse waveforms can be computed. This concludes our proposed method to for invariant-based state preparation in noisy dynamics. While the inverse-engineering steps addressing subproblem 1 in Sections \ref{subsec:traj}-\ref{subsec:thm}, are essential to ensure obtaining physical control pulses, different approaches compared to our proposal in \ref{sub2} could be utilized in order to address subproblem 2.

\section{Results}
In this paper, we perform two sets of numerical experiments to demonstrate the proposed method. The first set explores in detail the state preparation of a single-qubit subject to classical noise, which provides the foundation of our approach.  The second set shows examples of preparing higher-dimensional quantum states by restricting the dynamics to SU(2) as explained earlier. We show the cases of a qutrit subject to quantum noise as well as a noiseless two-qubit system.

\begin{figure}
    \centering
    \includegraphics[width=0.99\linewidth]{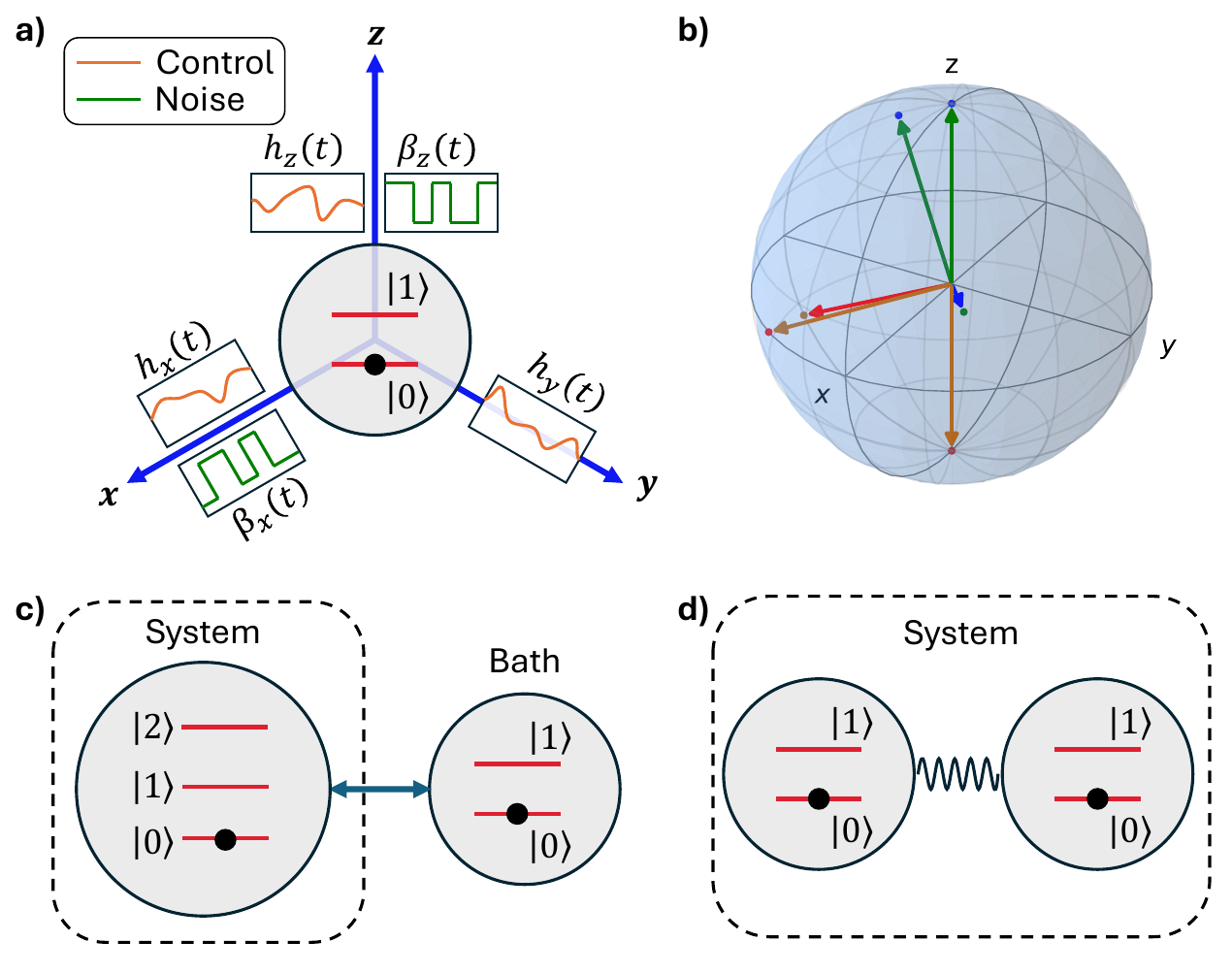}
    \caption{\textbf{System settings for the numerical simulations.} a) A qubit subject to RTN classical noise acting along X- and Z-axis, and control applied along all three axes. b) The Bloch sphere representation of the target states to be prepared for the single-qubit. The corresponding control Hamiltonians are shown in Table \ref{table:target}. c) A qutrit subject to quantum noise due to coupling to a two-level quantum bath, and assuming full SU(3) control. d) A two-qubit closed-system assuming full SU(4) control.}  
    \label{fig:setting}
\end{figure}

\subsection{Qubit state preparation}
\paragraph*{System setting}
In this paper, we numerically demonstrate the proposed method to control a qubit subject to multi-axis classical noise, thus exhibiting non-Markovian dynamics. The total Hamiltonian of the noisy system is given by
\begin{equation}
    H(t) = H_{\text{ctrl}}(t) + g_x\beta_x(t)\sigma_x + g_z\beta_z(t)\sigma_z
\end{equation}
where $\beta_x(t)$ and $\beta_z(t)$ are stochastic processes representing classical noise acting along $\sigma_x$ and $\sigma_z$, and $g_x$ and $g_z$ are the coupling strength between the system and the classical bath. Figure \ref{fig:setting}a depicts the system setting. Here, we choose $\beta_x(t)$ and $\beta_z(t)$ to be Random Telegraph Noise (RTN) processes with switching rate $\Gamma$ (See Supplementary Note 5 for formal definition). This noise is colored and non-Gaussian, (thus inducing non-Markovian qubit dynamics), and is observed in many physical systems, such as superconducting qubits \cite{rtnsq}. The system parameters are chosen as follows: $T=3.2 \ \mu s$, the evolution time interval is discretized into $4800$ time steps, $\Gamma=0.02$ MHz, coupling strength $g_z/\Gamma=13$ and $g_x/\Gamma=10$. In this paper, we use Dyson expansion up to the second order to simulate the noise dynamics \cite{W6}, as defined in Supplementary Note 2. We apply the proposed method to prepare the ground state of six different final control Hamiltonians, from the initial state $\ket{0}$ which is the ground state $H_{\text{ctrl}}(0)=\Omega\sigma_z$, where $\Omega=2\pi\times0.4$ Mrad/s. The final control Hamiltonians are randomly chosen and designed to cover all cases discussed in Section \ref{subsec:traj}. The target Hamitlonans are given in Table \ref{table:target}, and their ground states are plotted on the Bloch Sphere in Figure \ref{fig:setting}b. While Target (ii) and (vi) represent special cases of quantum memory and population inversion respectively, the other targets are random. 

\paragraph*{Implementation} 
The protocol was implemented in Python. The noise mitigation steps, including the whitebox and graybox, were implemented using tensorflow \cite{tf} and Keras \cite{keras} Python packages. The Bloch sphere plots were constructed using QuTiP Python package \cite{qutip}. 

\begin{table}[t]
\caption{\textbf{The final control Hamiltonians chosen in the numerical simulations of the qubit system}. The objective is to prepare the ground state of control Hamiltonians of the form $H_{\text{ctrl}}(T)=(h_x(T)\sigma_x + h_y(T)\sigma_y+h_z(T)\sigma_z)/2$ from the ground state of initial control Hamiltonian $H_{\text{ctrl}}(0)=-\Omega\sigma_z/2$, in the presence of noise. The examples chosen cover all cases discussed in Section \ref{subsec:traj}. }
\label{table:target}
\centering
\begin{tabular}{|c|c|c|c|c|}
\hline 
\textbf{ID} &
{$\mathbf{h_x(T)}$} & $\mathbf{h_y(T)}$ & $\mathbf{h_z(T)}$ & \textbf{Case}\\
\hline
(i) & $\frac{\Omega}{2}$ & $\frac{\Omega}{\sqrt{3}}$ & $-\Omega$ & 1\\
(ii) & $0$ & $0$ & $-\Omega$ & 1\\
(iii) & $\Omega\sqrt{\frac{5}{7}}$ & $\Omega\sqrt{\frac{2}{7}}$ & $0$ & 2\\
(iv) & $\frac{\Omega}{\sqrt{2}}$ & $\frac{\Omega}{3}$ & $-\Omega\sqrt{\frac{7}{18}}$ & 2\\
(v) & $0$ & $\Omega\sqrt{\frac{4}{5}}$ & $\frac{\Omega}{\sqrt{5}}$ & 3\\
(vi) & $0$ & $0$ & $\Omega$ & 4\\
\hline
\end{tabular}
\end{table}

\begin{figure*}
    \centering
    \includegraphics[width=\linewidth]{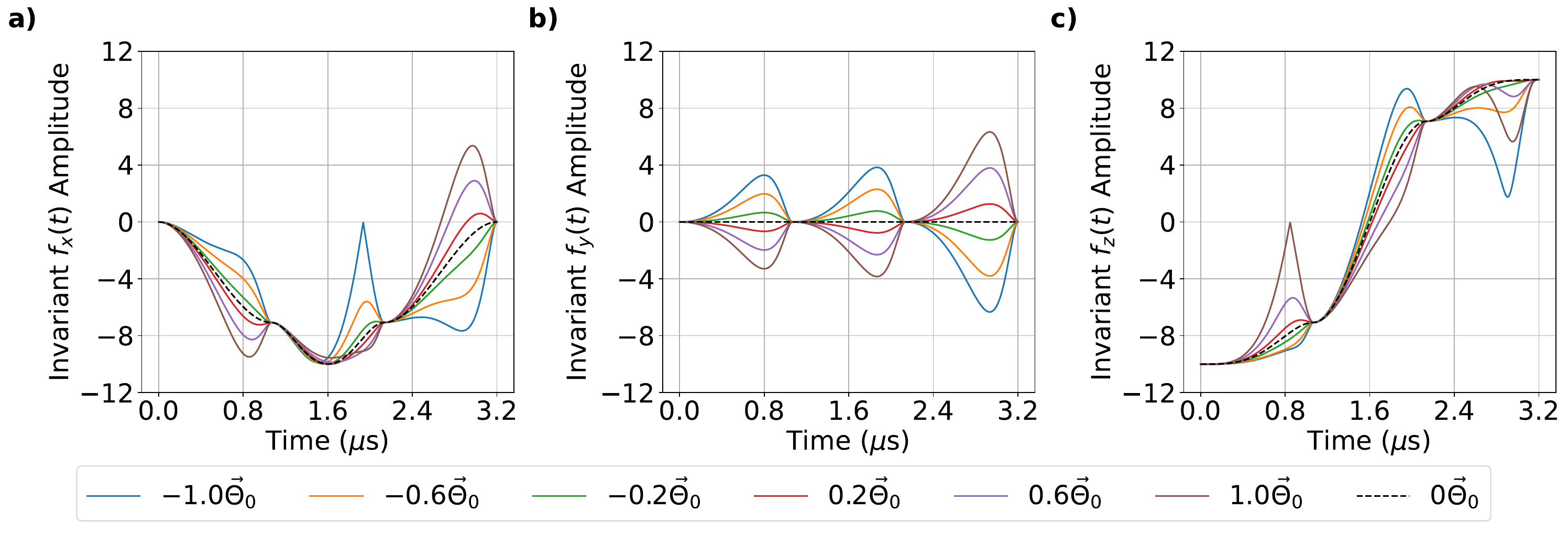}
    \caption{\textbf{Family of invariants for target (vi) for the qubit system.} This case requires 3 subtrajectories. Different examples from the invariant family are plotted as function of time. The parameters are chosen to be of linear scaling of $\vec{\Theta}_0$ that corresponds to the extreme case where $f_3(t)\to 0$ at exactly 1 point. The dotted curve represents degree 3 solution, i.e. the minimal polynomial that satisfies the boundary conditions. }
    \label{fig:family}
\end{figure*}

\paragraph*{Invariant Design}
As presented in Section \ref{subsec:traj}, we design the subtrajectories based on the target state. Particularly, target (i) and (ii) do not require subdividing the evolution trajectory. Targets (iii), (iv) and (v) require two subtrajectories. And finally, target (vi) requires three subtrajectories. The reference axes are set to be $Z$, $Z$, $Z\to X$, $Z\to X$, $Z\to Y$, and $Z\to X \to Z$ for each of the target states respectively. We then find the boundary invariants for each subtrajectory following Section \ref{sec:boundary}. Next, we choose the polynomials representing the invariant coefficients $\tilde{f}_1(s)$ and $\tilde{f}_2(s)$, as in Section \ref{subsec:invdgn}, to be of degree 18. This gives a total of $2\times 15=30$ free parameters per subtrajectory. After that, we calculate the bound of the free variables $v_{\max}$ for each subtrajectory according to Section \ref{subsec:vmax}, ensuring real non-vanishing $\tilde{f}_3(s)$, and thus singularity-free control pulses as shown in Section \ref{subsec:thm}. In Supplementary Table S1, we show the computed values of $v_{\max}$ in each subtrajectory, for all target states. In the cases of more than 1 subtrajectory, we choose to have the normalized free variables $v_{k,j}/v_{\max}$ to be the same across all subtrajectories, to reduce the complexity of the optimization problem. For instance, for target (vi), we have total of 30 independent free parameters instead of 90. Figure \ref{fig:family} shows a subset of the family of invariants for target (vi). In this Figure, we find the extreme-case scenario $\vec{\Theta}_0$, where $f_3(t)=0$ at exactly 1 point. We then scale it linearly by different factors and plot the overall invariant as a function of time. Supplementary Figures (S2-S6) show similar plots for the other targets. We can then obtain the control pulses directly from any given member of the invariant family. Note, while we chose an extreme case to show in these figures (where $f_3(t)=0$ at one point), for dataset generation and optimization procedure, we enforce the free parameters to be slightly less than $v_{\max}$ to completely avoid this case. Particularly, we set the actual bound on the free variables to be $(1-\epsilon)v_{\max}$, where $\epsilon=10^{-5}$.
\paragraph*{Dataset generation} For each target state, we construct a dataset of 10,000 control pulses sampled randomly from the pulse family, and compute the corresponding Pauli matrix observables for the noisy qubit at the end of the evolution. In this paper, we use Dyson expansions truncated to second-order in the coupling strength for simulating the system dynamics (See Supplementary Note 5 for detailed computations). The dataset serves two purposes. The first is to provide a benchmark for the performance of different pulses. This is done by computing the fidelity for each pulse using Equation \ref{eq:fidelity}, and then plotting the histogram of those fidelities. The second purpose is to train the graybox model for the noise mitigation step. The randomization of the control pulses is done by sampling the free variables of the invariant in $\vec{\Theta}$, maintaining that they are bounded in the interval $[-v_{\text{max}}, v_{\text{max}}]$ in each subtrajectory. Empirically, we found that using a skewed distribution that favors parameters near $v_{\text{max}}$, such as Beta distribution, results in a wider range of pulse shapes compared to a uniform distribution. Thus, this gives better opportunity of finding optimal pulses to cancel the noise effects. Here, we use Beta distribution with shape parameters $\alpha=\beta=0.1$.

\paragraph*{Optimal control assuming known noise}
Following Section \ref{sub2}, we minimize the cost function to find the optimal pulse parameters for each target state using the search algorithm in Supplementary Note 4. Since, this optimization requires computing the Dyson terms, which is computationally intensive, we initialize the search algorithm with one initial point (corresponding to vanishing free parameters), and perform 100 iterations to find the optimal control pulse. 

\paragraph*{Optimal control assuming unknown noise}
In this case, we use the machine learning approach described in Section \ref{sub2}. The detailed architecture of our proposed graybox is given in Supplementary Note 6. The first step is to train the graybox model over the dataset of each target state. We split the dataset into training and testing sets with ratio of 80:20 respectively. We use Adam optimizer \cite{Adam} with learning rate $5\times 10^{-4}$ and perform 1000 iterations for the training. We validated the training performance using the testing data, and to ensure no overfitting occurs. The trained model is then used in the optimization algorithm to find the optimal control pulse parameters. The graybox model is computationally efficient. This is due to the fact that the blackbox part of the model learns a direct mapping from parameters $\vec{\Theta}$ to $V_O(T;\vec{\Theta})$, which is more efficient that computing the Dyson terms in the whitebox approach. Thus, we start the random search with 100,000 initial random values for $\vec{\Theta}$ sampled using the Beta distribution, and select the best one to run the iterations. The optimization algorithm is performed for a 100 iterations.

\begin{figure*}
    \centering
    \includegraphics[width=\linewidth]{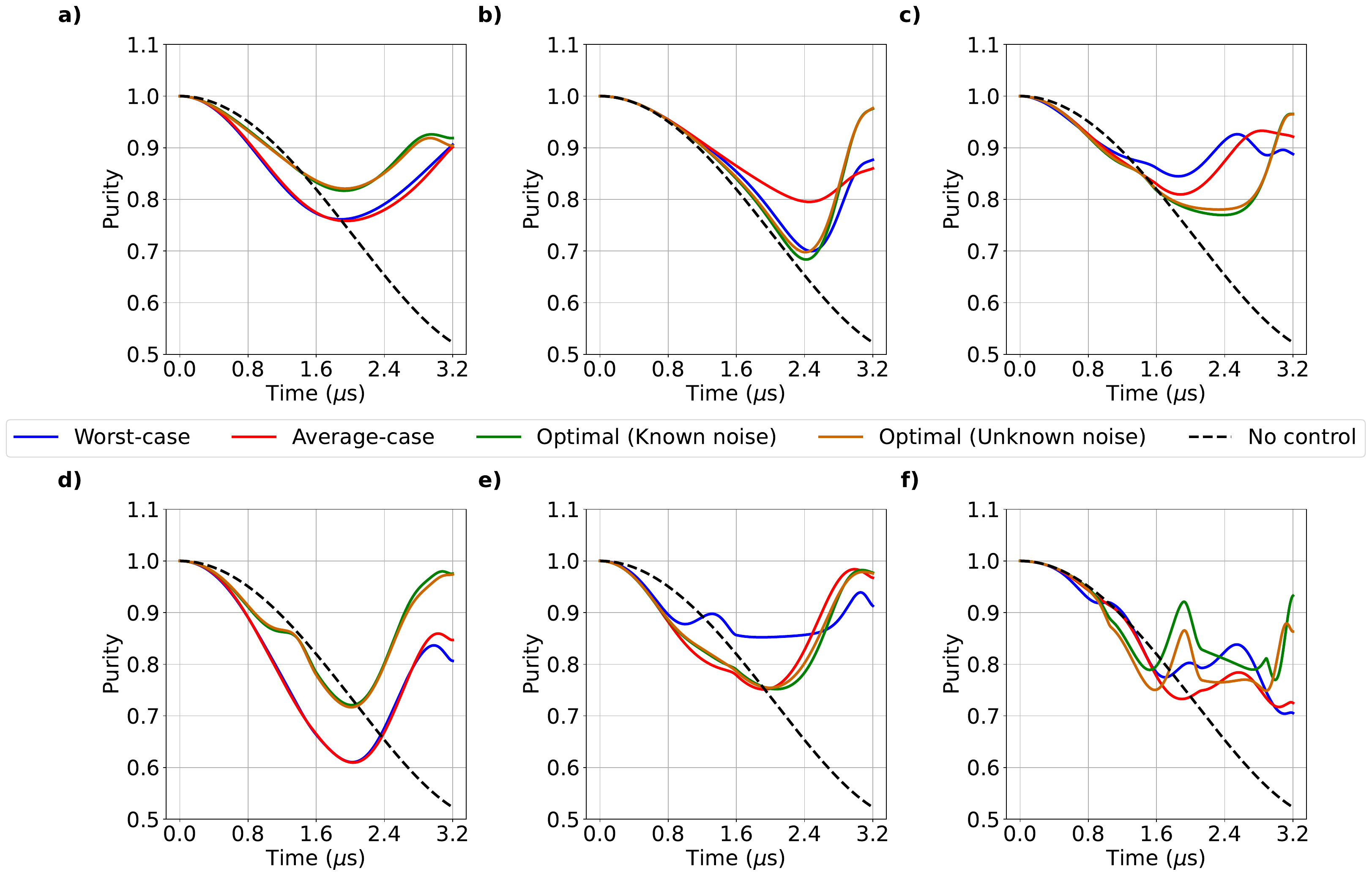}
    \caption{\textbf{State purity of the qubit system for different pulses sequences.} The purity of the state is plotted as a function of time for the worst-case and average-case pulse from the dataset, and the optimized pulse (known and unknown noise), in comparison to the case where the system is subject purely to the noise (i.e. $H_{\text{ctrl}}(t)=0$).
    Plots a)-f) correspond to the targets (i)-(vi), respectively.}
    \label{fig:purity}
\end{figure*}

\begin{figure*}
    \centering
    \includegraphics[width=\linewidth]{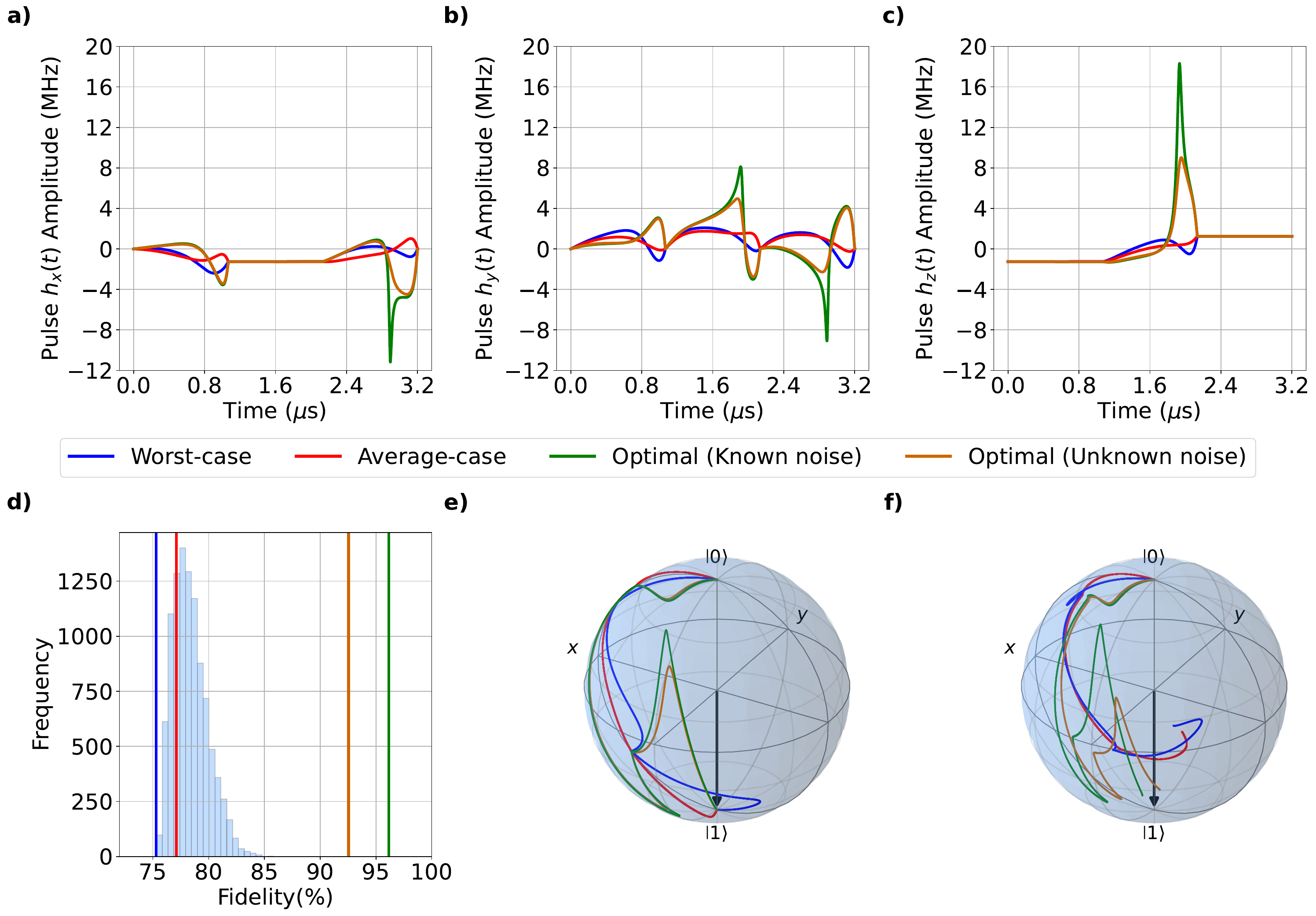}
    \caption{\textbf{Optimization results for target (vi) for the qubit system.} We compare the optimization result (known and unknown noise) with the worst-case and average-case pulse in from the dataset. The optimal pulse in known noise setting is obtained using a whitebox approach, while graybox approach is used to find optimal pulse for unknown noise setting. a), b) and c) show the pulse waveforms as function of time along x, y and z-axis, respectively. d) shows the fidelity of these pulses, in relation to the histogram of fidelities computed over the whole dataset. e) shows the state trajectory generated by each of these pulses on the Bloch sphere for a closed system, while f) shows the open system trajectory.}
    \label{fig:result_main}
\end{figure*}

\paragraph*{Optimization results} We present the control performance under known and unknown noise in Table \ref{table:optimization}. The final state fidelity is also shown for worst-case, average-case, and best-case pulse in the dataset for each target, along with the optimal pulse obtained from whitebox and graybox approaches. Figure \ref{fig:purity} shows a plot of the purity of the state of the system over time for each of the 6 targets. The plot is for the cases of the worst-case pulse, average-case pulse, and optimal pulse in both known and unknown noise settings, in comparison with no control (i.e. only noise affecting the system dynamics). In Figure \ref{fig:result_main}, we show a comparison between worst-case pulse, average-case pulse, and optimal pulse in both known and unknown noise settings, for target (vi). Figure \ref{fig:result_main}a-Figure \ref{fig:result_main}c show the pulse waveform $h_x(t)$, $h_y(t)$, and $h_z(t)$. Figure \ref{fig:result_main}d shows the histogram of the fidelities over the whole dataset in relation to the four aforementioned control sequences. \ref{fig:result_main}e show the intermediate dynamics of the each control sequence depicted as trajectories on the Bloch sphere, in the absence of noise. Finally, in \ref{fig:result_main}f we show similar plot in the presence of noise. Supplementary Figures (S7-S11) show similar plots for the other five targets. 

\subsection{Qudit state preparation}
\begin{figure*}
    \centering
    \includegraphics[width=0.7\linewidth]{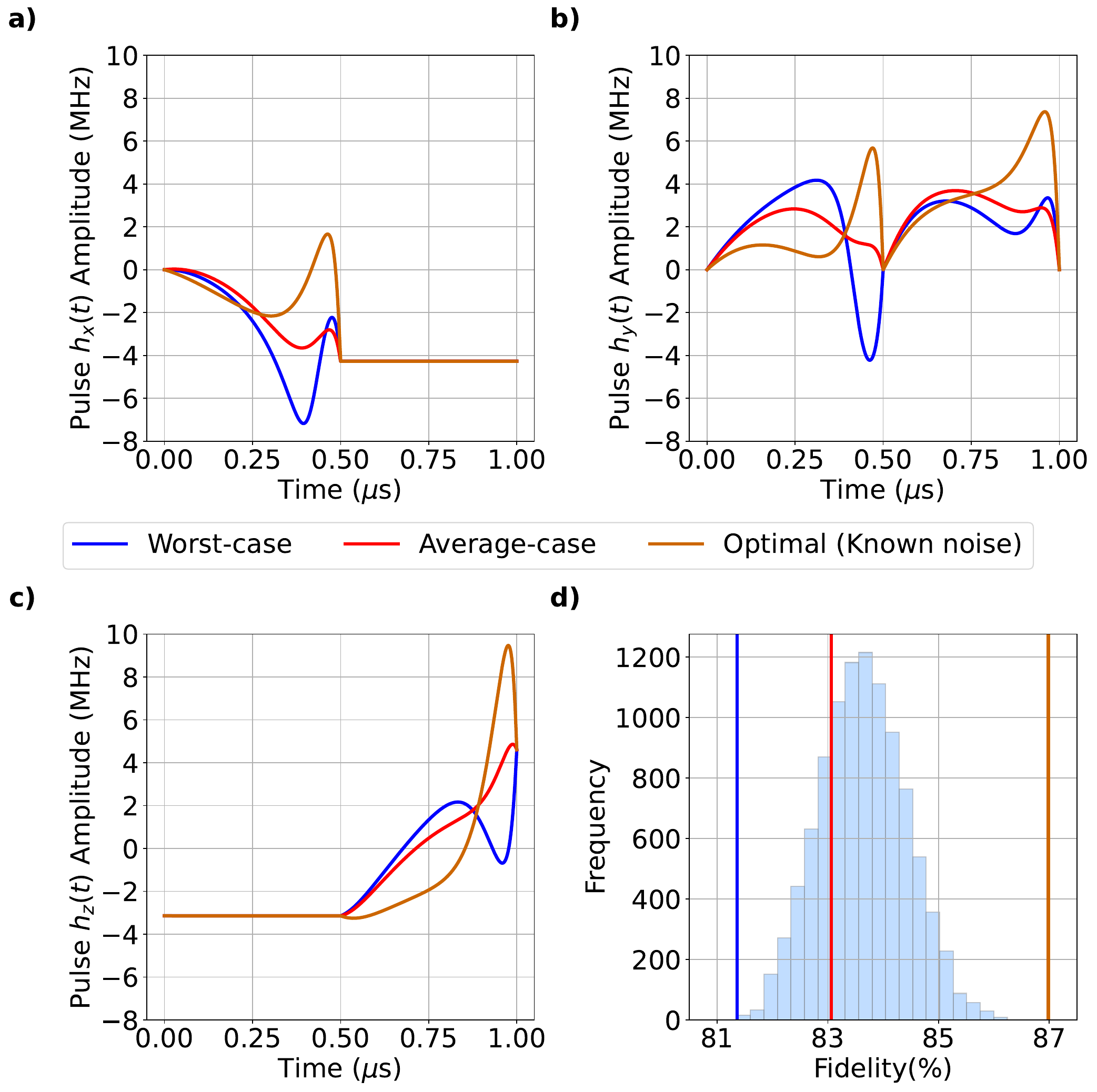}
    \caption{\textbf{Optimization results for state preparation of a qutrit in the presence of quantum noise.} We compare the optimization result (known noise) with the worst-case and average-case from the dataset. The optimal pulse in known noise setting is obtained using the whitebox approach. a), b) and c) show the pulse waveforms as function of time along $\Sigma_x^{(3)}$, $\Sigma_y^{(3)}$ and $\Sigma_z^{(3)}$ bases, respectively. d) shows the fidelity of these pulses, in relation to the histogram of fidelities computed over the whole dataset}.
    \label{fig:result_q3}
\end{figure*}

Next, we show two examples of higher-dimensional state preparation. The first example is for a qutrit subject to quantum noise. Here, we model the environment as a qubit that is coupled to the qutrit. The total Hamiltonian is given by
\begin{align}
H(t) &= H_{\text{ctrl}}(t) + H_\text{B}(t) + H_{\text{SB}}(t),
\end{align}
where
\begin{align}
H_{\text{ctrl}}(t) &= \frac{1}{2}(h_x(t)\Sigma_x^{(3)} +h_y(t)\Sigma_y^{(3)} + h_z(t)\Sigma_z^{(3)} ) \\
H_B(t) &= \omega\sigma_z \\
H_{SB}(t) &= A(S_x\sigma_x + S_y\sigma_y + S_z\sigma_z).  
\end{align}
The $3\times3$ matrices $\Sigma_x^{(3)}, \Sigma_y^{(3)}, \Sigma_Z^{(3)}$ are defined based on the initial and target states as will be shown, and 
\begin{align}
    S_x &= \frac{1}{\sqrt{2}}\begin{pmatrix}
        0 & 1 & 0\\
        1 & 0 & 1\\
        0 & 1 & 0
    \end{pmatrix}\\
    S_y &= \frac{1}{\sqrt{2}}\begin{pmatrix}
        0 & -i & 0\\
        i & 0 & -i\\
        0 & i & 0
    \end{pmatrix}\\
    S_z &= \begin{pmatrix}
        1 & 0 & 0\\
        0 & 0 & 0\\
        0 & 0 & -1
    \end{pmatrix}.
\end{align}
This coupling term in the Hamiltonian causes quantum noise in the qutrit. Figure \ref{fig:setting}c depicts this model. The system parameters are chosen as follows: $T=1 \ \mu s$, the evolution time interval is discretized into $4800$ time steps, $\Omega=2\pi\times0.4$ Mrad/s, $\omega=2\pi\times 0.2$ Mrad/s and coupling strength $A=2\pi\times 0.07$ Mrad/s. Simulation of the system dynamics was performed by time evolving the joint state of qutrit and qubit, followed by tracing out of the qubit. We apply the proposed method to prepare the randomly-chosen target state
\begin{align}
    \ket{\psi(T)} &= \sqrt{\frac{2}{15}}\ket{0}+e^{1.1i}\sqrt{\frac{5}{24}}\ket{1}+e^{0.4i}\sqrt{\frac{79}{120}}\ket{2},
\end{align}
from the initial state $\ket{0}$. In the SU(2) picture, the basis can be computed to be
\begin{align}
    \ket{0_S} &= \ket{0} \\
    \ket{1_S} &= \sqrt{\frac{15}{13}}\left(e^{1.1i}\sqrt{\frac{5}{24}}\ket{1}+e^{0.4i}\sqrt{\frac{79}{120}}\ket{2}\right),
\end{align}
which we can use to compute each of the $\Sigma_x^{(3)}, \Sigma_y^{(3)}, \Sigma_z^{(3)}$ matrices. This setting corresponds to case 2 of the trajectory splitting, which would require construction of two subtrajectories. A dataset of 10,000 random examples was constructed to plot the fidelity histogram. The optimal pulses were obtained using a whitebox model. We present the performance of our control protocol for a qutrit in Figure \ref{fig:result_q3}, where we show a comparison between worst-case pulse, average-case pulse, and optimal pulse.

The second example is a two-qubit system which is assumed to be noiseless, i.e., 
\begin{align}
H(t) &= H_{\text{ctrl}}(t) \\
 &= \frac{1}{2}(h_x(t)\Sigma_x^{(4)} +h_y(t)\Sigma_y^{(4)} + h_z(t)\Sigma_z^{(4)} ). 
\end{align}
Figure \ref{fig:setting}d depicts this case. The system parameters are chosen as: $T=1 \ \mu s$ discretized into $4800$ steps, and $\Omega=2\pi \times 0.4$ Mrad/s. We are interested in preparing the fourth Bell state
\begin{align}
    \ket{\psi_T} = \ket{\Psi^-} = \frac{1}{\sqrt{2}}(\ket{01}-\ket{10}).
\end{align}
In this case, we can compute the basis 
\begin{align}
    \ket{0_S}&=\ket{00}\\
    \ket{1_S}&= \frac{1}{\sqrt{2}}(\ket{01}-\ket{10})
\end{align}
and the corresponding $\Sigma_x^{(4)}, \Sigma_y^{(4)}, \Sigma_z^{(4)}$ matrices. This corresponds to case 4 in the SU(2) picture and would require 3 subtrajectories. The pulse design for the subspace is performed exactly the same as that for target (vi) of the single qubit. Figure \ref{fig:fidelity_2q} shows the evolution of the fidelity between the target state and the two-qubit state for a subset of control pulses in the family. This subset is chosen by finding the extreme-case scenario $\Theta_0$ where $f_3(t)=0$ at exactly one point, and taking a linear scaling of this $\Theta_0$ between -1 to 1. Supplementary Figure S12 shows the pulse waveform in the complete Pauli basis for each of these pulses.

\begin{table}[t]
\caption{\textbf{Fidelity(\%) of optimal control pulses for the qubit system.} We compare the fidelity (expressed as percentage) of the optimal control pulses obtained through whitebox (WB) and graybox (GB) approaches against the minimum, average, and maximum fidelity observed in the dataset, for each target.}
\label{table:optimization}
\centering
\begin{tabular}{|c|c|c|c|c|c|}
\hline
\textbf{ID} & \textbf{Min} & \textbf{Avg} & \textbf{Max} & \textbf{WB} & \textbf{GB} \\
\hline
(i) & 89.16 & 90.41 & 94.36 & 95.07 & 94.40\\
(ii) & 89.24 & 91.10  & 93.41  & 96.48 & 96.30\\
(iii) & 93.85 & 95.28 & 97.68 & 98.03 & 97.95\\
(iv) & 83.29 & 87.41 & 95.71 & 97.05 & 96.85\\
(v) & 94.06 & 97.62 & 97.88 & 97.93 & 97.86\\
(vi) & 75.30 & 77.11 & 85.83  & 96.14 & 92.54\\
\hline
\end{tabular}
\end{table}

\section{Discussion and future work}
In this paper, we have introduced a method for preparing finite-dimensional states in the presence of non-Markovian open system dynamics using dynamical invariants. By splitting the target evolution trajectory into subtrajectories based on initial and final control Hamiltonians, and with proper parameterization, the control pulses are guaranteed to be singularity-free at all time steps. This has been major issue with previous works adopting invariant-based control methods. We also addressed the noise mitigation subproblem through model-based control, either whitebox or graybox depending on the noise assumptions. Finally, we introduce a strategy for extending this protocol to finite-dimensional systems, by constructing an SU(2) subspace based on the initial and target states. The novelty of the present work is therefore the combination of dynamical-invariant pulse design, trajectory segmentation, bounded parameterization, and extension to finite-dimensional systems, together with an optimization layer for robustness under non-Markovian noise.

The SU(2) subspace construction step achieves significant improvement in the computational complexity of the pulse engineering procedure. The invariant computations remain in SU(2) regardless of the original qudit dimensionality. However, this works under the assumption of having full control over a complete orthonormal basis set. Experimentally, this might be challenging as many systems will only have access to a subset of independent controls.

Our method for avoiding the singularity in pulse design is based on splitting the evolution into subtrajectories ensuring the continuity of the reconstructed pulses at all times, as shown in Section \ref{sec:boundary}.
The pulse waveforms shown in Figures \ref{fig:result_main} and \ref{fig:result_q3} demonstrate this fact. This avoids additional bandwidth requirements associated with piecewise-discontinuous protocols. The switching of the reference axis happens at the pulse design stage and does not translate to any additional cost in the physical implementation. In other words, the pulse waveforms over the full evolution time are obtained, then applied in experiment without further changes.

Moreover, the design of our protocol yields an infinite family of control pulses, from which the optimal one can be selected according to various criteria, such as overlap with noise, leakage into higher levels in multi-dimensional systems, or control pulse energy. Moreover, most quantum control methods rely on inverting the Schr\"odinger equation to determine the optimal Hamiltonian that achieves a desired target state. This constitutes an inverse partial differential equation problem, which is generally computationally hard to solve. However, the invariant-based approach transforms this analytic tasks into a much simpler algebraic one--namely, fitting the functional form of the invariant to satisfy a prescribed set of boundary conditions. With an appropriate parameterization, this fitting can be made exact rather approximate, yielding a globally optimal solution to the closed-system control problem, something that is generally hard to guarantee for most numerical control methods. Although the works on quantum control landscapes \cite{rabitz_landscape_2004,qcland} show that gradient-based searches can be efficient and need not be hindered by suboptimal traps for any unconstrained controllable system, in practice this is challenging. As shown in \cite {riviello_constraints_2015}, additional constraints on pulse amplitude, bandwidth, duration, or parametrization can make the optimization more difficult and prone to falling into local traps. Furthermore, the GRAPE method \cite{grape} and its variants, a widely adopted set of algorithms, optimize the control pulse at each instant in time using gradient-based descent schemes, which are susceptible to convergence toward local minima. Thus, they often require careful initialization and can be computationally expensive for high-dimensional systems or when modeling non-Markovian noise. Other approaches employing efficient parameterization of control pulses, such as frequency domain encoding \cite{crab1}, still lack guarantees of global optimality, even for closed systems. In contrast, our framework directly constructs families of admissible pulses through algebraic relations, ensuring that boundedness and boundary conditions are enforced analytically. Thus, we do not require any optimization procedure in the closed-system setting, avoiding the aforementioned challenges. Additionally, our approach is substantially more computationally efficient, as the entire quantum trajectory can be parametrized with only a small number of free variables. The computational complexity of our method is independent of the number of discretized time steps and is mathematically guaranteed to yield optimal solutions in the closed-system regime. Furthermore, the integration with ML modules provide adaptability to complex, partially unknown noise environments. These features collectively make the proposed approach more robust, scalable, and transparent than GRAPE for realistic, noisy quantum platforms.

Another attractive feature of dynamical invariants is providing an efficient means of computing the system's state at any given time $t=\tau$ by evaluating only the eigenvector of the invariant operator $I(\tau)$ at that instant. This approach incurs significantly lower computational complexity compared to approximating the time-ordered evolution, which requires evolving from $t=0$ up to $t=\tau$ over a large number of discrete time steps to achieve sufficient accuracy, and consequently performing a correspondingly large number of matrix multiplications to evolve from $U_{\text{ctrl}}(0)$ to $U_{\text{ctrl}}(\tau)$.

Regarding the noise mitigation step, the combination of the noise operator formalism with machine learning enables modeling of arbitrary non-Markovian dynamics without relying on simplifying assumptions or approximations, while rigorously preserving physical constraints such as state positivity. This framework allows reliable state preparation even when the noise affecting the system is unknown or its explicit model is computationally intractable. It is worth noting that the invariant-based construction does not search the full control landscape. Accordingly, we do not claim global optimality over all admissible pulse shapes in the open-system setting. Rather, the advantage of the method is that it restricts the search to a structured family of bounded pulses that exactly solve the closed-system boundary-value problem and can then be optimized according to additional criteria such as robustness against noise.

The graybox structure introduced in this work exhibits improved computational complexity compared to previous approaches \cite{W1,W2,W3,W4,W5,W6,gbyule}, as it requires only a subset of informationally complete set of measurements, thereby simplifying both dataset generation and model training. Furthermore, the use of dynamical invariants provides exact knowledge of the final state in the absence of noise, eliminating the need for the whitebox layer that computes the time-ordered evolution of control pulses used in prior studies. This feature offers a significant advantage for higher-dimensional and can be further extended to continuous-variable systems. An important thing to note is that the graybox approach does not require a Markov approximation, but it does assume that the noise statistics remain sufficiently stable over the period in which the training data are collected and the model is used. If the device operates in a strongly non-stationary regime with rapid parameter drift, the learned model would need to be refreshed through recalibration, online adaptation, or retraining. The corresponding experimental cost is similar in spirit to other data-driven calibration procedures: the dataset is accumulated over repeated runs and therefore is not limited by the coherence time of a single realization, although excessive drift during data acquisition would naturally reduce the reliability of the learned model. 

Figure \ref{fig:family} and Supplementary Figures (S2-S6) illustrate a broad variety of invariant waveforms obtained, which in turn give rise to a wide range of corresponding control sequences. This diversity can enhance the protocol's noise mitigation performance by providing greater flexibility in pulse shaping compared to approaches employing a more restricted set of control pulses. We also note that all invariant waveforms presented are continuous throughout the entire evolution time and converge smoothly at the boundaries of each subtrajectory.

As shown in Table~\ref{table:optimization}, the whitebox optimization, used in the known-noise setting, consistently outperforms the best-case pulse observed in the dataset. The corresponding fidelities are higher by 0.72\%, 3.06\%, 0.38\%, 1.35\%, 0.05\%, and 10.31\% for each target, respectively. The graybox optimization, applied in the unknown noise setting, similarly achieves high fidelities, surpassing the best-case pulse in the dataset for nearly all targets. With the exception of target~(v), whose fidelity is 0.02\% lower than the best-case but 0.2\% higher than the average-case, the observed fidelities exceed the best-case values by 0.05\%, 2.99\%, 0.27\%, 1.14\%, and 6.71\% for the remaining targets. Overall, the results indicate that graybox optimization achieves performance comparable to, and only marginally below, that of the whitebox method. This demonstrates the effectiveness of the machine-learning-based approach in modeling noise without any prior knowledge or assumptions. It is also important to note that the whitebox model used here is perfectly matched to the simulation model, a condition rarely met in practical settings. In experimental systems, deviations from idealized models commonly arise due to fabrication imperfections and stochastic noise environments. Consequently, a machine learning based approach offers a more robust and practical solution in such scenarios.

Figure~\ref{fig:purity} shows that the application of control pulses enhances the qubit lifetime. This is evident from the behavior of the state purity, which approaches its minimum value of $1/2$ toward the end of the evolution in the absence of control. In contrast, the state purity remains significantly higher when control pulses are applied. For example, the purity for target~(i) is observed to be $0.9$ or higher at the end of the evolution. At intermediate times, the purity exhibits pronounced variations due to the action of the control fields. The worst-case pulse produces the final state with the lowest purity among the four cases, except for targets~(i) and~(ii). This can be understood by recalling that pure and mixed states can exhibit the same fidelity with respect to a target state. For instance, the pure state $\ket{\psi_1} = \sqrt{0.9}\ket{0} + \sqrt{0.1}\ket{1}$ and the mixed state $\rho_2 = (\mathbb{I} + 0.8\sigma_z)/2$ yield the same fidelity with $\ket{0}$. This observation suggests the potential benefit of including purity as part of the cost function. Finally, we note that the whitebox and graybox optimal pulses generally yield higher final-state purity compared to the worst-case and average-case pulses.

Figure~\ref{fig:result_main} and Supplementary Figures~(S7-S11) show clear differences in the pulse shapes across all control directions for the optimal pulses compared to the worst-case and average-case scenarios. The fidelities achieved by the optimal pulses are generally higher than the distribution observed in the dataset histograms. Under closed-system conditions, the four pulses yield distinct intermediate dynamics yet converge to the same target state. In the presence of noise, however, the optimal pulses produce a final state that remains significantly closer to the target than in the other two cases, consistent with the fidelities reported in Table~\ref{table:optimization}. For all target states, the optimal controls obtained from the graybox and whitebox approaches generate similar dynamics and closely matching final states, highlighting the effectiveness of the graybox model. Overall, these results demonstrate that our protocol provides a comprehensive and promising framework for controlling non-Markovian dynamics.

\begin{figure}
    \centering
    \includegraphics[width=0.9\linewidth]{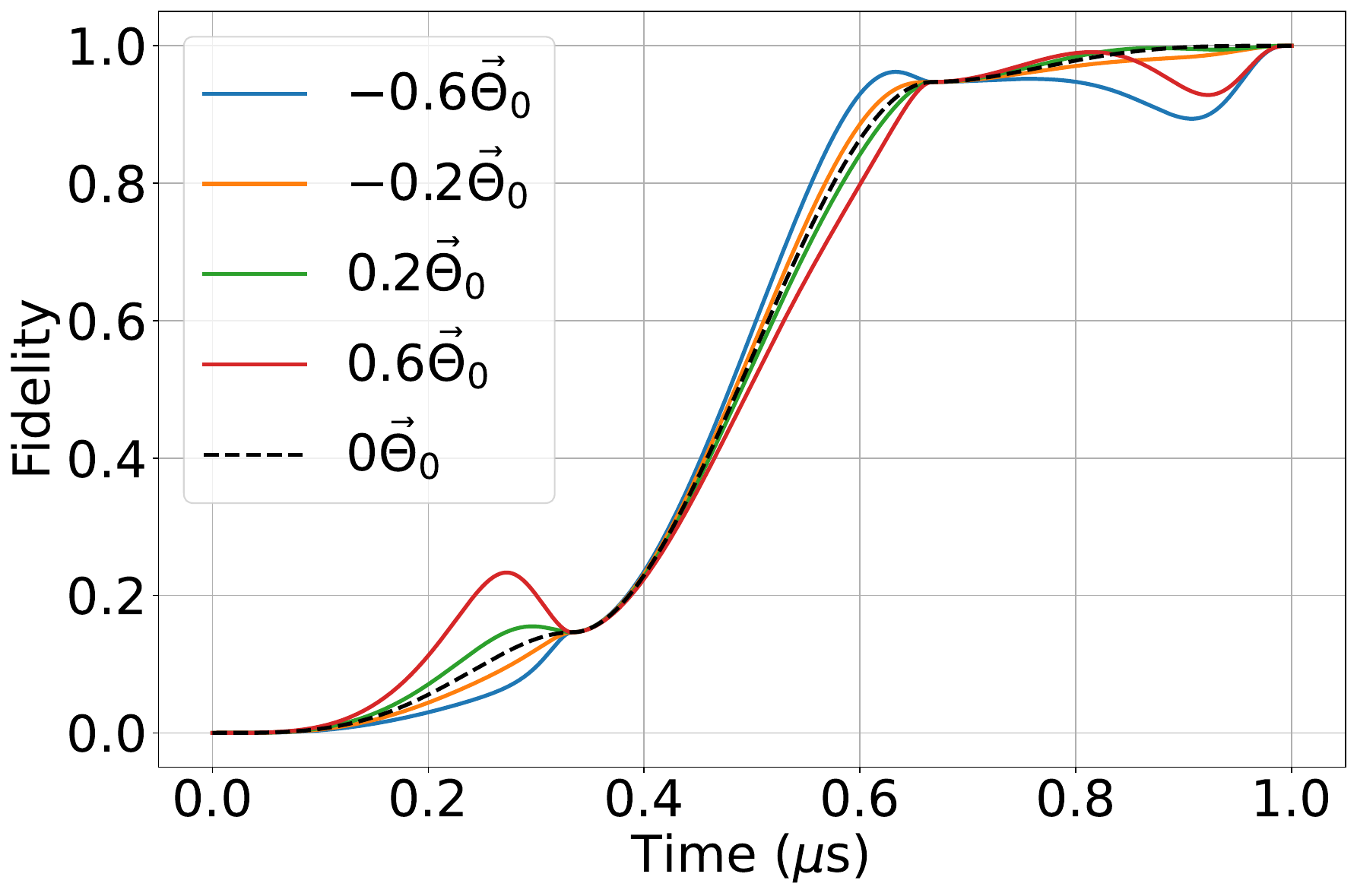}
    \caption{\textbf{State fidelity evolution over time for a family of pulses that prepares the $\ket{\Psi^-}$ 2-qubit EPR state in the absence of noise. } The examples correspond to a linear scaling of $\vec{\Theta}_0$, which defines an extreme-case where $f_3\to 0$ at exactly one point. The dotted line shows the baseline case $\vec{\Theta}_0 = 0$.}
    \label{fig:fidelity_2q}
\end{figure}

Figure \ref{fig:result_q3} shows that the worst fidelity observed in the dataset is 81.36\% and the average fidelity is 83.07\%, while the optimal pulse shows 86.98\% fidelity. For the two-qubit closed-system, the fidelity of instantaneous states follow different trajectories for different pulses in the family, but converge at the final evolution time, as shown in Figure \ref{fig:fidelity_2q}. This shows that the proposed protocol is successful in designing control pulses, even in higher-dimensional systems. This could be integrated with a suitable noise mitigation optimal control method, as shown in other example systems.

The requirement for full basis control could be addressed in future by selecting an SU(2) subspace that lies within the experimentally accessible subspace.
Another possible direction is to construct a subspace which is not SU(2) but provides experimentally available control pulses while reducing the computation complexity as compared to addressing the full qudit using invariants. A different approach could be starting a different Lie algebraic structure that can encode the  constrained Hamiltonian, and then obtaining the corresponding invariant parametrization. Another extension to this work is to explore its applicability to continuous-variable systems. Some attempts have previously been made to control these systems using dynamical invariants~\cite{levy2018noise,iontrap,iontrap2} in the noiseless regime based on utilizing Lie algebraic methods. Further extensions include the implementation of quantum gates, which would require controlling the global phase shift of the invariant eigenstates, as described by equation \ref{eq:phase}. This can been performed, for example, by designing the parametrization that has vanishing dynamical phase as shown in \cite{invcat}. Finally, we could investigate alternative cost functions and optimization strategies for enhanced noise mitigation. \\

\section{Data Availability}
The data used in this study is publicly available at \url{https://github.com/Ritik-sareen/DIPE}.

\section{Code Availability}
The source code for generating the invariants and control pulses is publicly available at \url{https://github.com/Ritik-sareen/DIPE}

\section{Acknowledgements} RS, AY and AP disclose support for the research of this work from the Australian Government through the Australian Research Council under the Centre of Excellence scheme (No: CE170100012). AP acknowledges an RMIT University Vice-Chancellor’s Senior Research Fellowship and a Google Faculty Research Award. This research was also undertaken with the assistance of resources from the National Computational Infrastructure (NCI Australia), an NCRIS enabled capability supported by the Australian Government. 

\section{Author Contribution}
RS designed and implemented the protocol, and conducted all the numerical experiments presented in this paper, with feedback from AY and AP. AY contributed to the implementation of the whitebox and graybox machine learning models. RS and AY wrote the paper with feedback from AP. AY and AP supervised the project.

\section{Competing Interests}
The authors declare no competing financial or non-financial interests.

\bibliography{bib}
\bibliographystyle{unsrt}

\end{document}

\title{Supplementary Materials for Singularity-free dynamical invariants-based quantum control}
\author{Ritik Sareen}
\address{Quantum Photonics Laboratory and Centre for Quantum Computation and Communication Technology, RMIT University, Melbourne, VIC 3000, Australia}
\address{School of Electrical Engineering and Telecommunications, UNSW Sydney, Sydney, NSW 2052, Australia}

\author{Akram Youssry}
\address{Quantum Photonics Laboratory and Centre for Quantum Computation and Communication Technology, RMIT University, Melbourne, VIC 3000, Australia}
\address{School of Electrical Engineering and Telecommunications, UNSW Sydney, Sydney, NSW 2052, Australia}

\author{Alberto Peruzzo}
\address{Quantum Photonics Laboratory and Centre for Quantum Computation and Communication Technology, RMIT University, Melbourne, VIC 3000, Australia}
\address{Quandela, Massy, France}

\maketitle

\section*{Supplementary Note 1: Finding a control Hamiltonian given its ground state}
It is required to find a control Hamiltonian $H_{\text{ctrl}}^{\psi}$ such that a given $\ket{\psi}$ is its ground state, i.e.,
\begin{equation}
    H_{\text{ctrl}}^{\psi}\ket{\psi} = \mu_{g}\ket{\psi},
\end{equation}
where $\mu_g$ is the minimum eigenvalue of the control Hamiltonian. To address this problem, we need to find a set of $d$ orthonormal states that includes $\ket{\psi}$, where $d$ is dimension of the Hilbert space. These states would act as the eigenstates of the Hamiltonian. The first step is to find the nullspace of $\bra{\psi}$, which can be computed using Singular Value Decomposition (SVD). The SVD of $\bra{\psi}$ is given by 
\begin{equation}
    \bra{\psi} = USV^\dagger,
\end{equation}
with the basis of nullspace of $\bra{\psi}$ given by the columns of $V$, which are orthonormal. Since $\text{rank}(\bra{\psi})=1$, then  $\text{nullity}(\bra{\psi})=d-1$ from the rank-nullity theorem. In other words, the number of columns of $V$ is $d-1$. Moreover, from the definition of nullspace, all the vectors spanned by the columns of $V$ are orthogonal to $\bra{\psi}$. We can then construct a unitary matrix $Q$ by augmenting $\ket{\psi}$ with the columns of $V$, i.e. $Q = \begin{pmatrix}
    \ket{\psi} & V
\end{pmatrix}$. This matrix would be considered as the matrix of eigenvectors of $H_{\text{ctrl}}^{\psi}$. Next, we construct a diagonal matrix $\Lambda$, representing the matrix of eigenvalues of $H_{\text{ctrl}}^{\psi}$. We choose the elements of $\Lambda$, to be distinct real numbers ranging from -1 to 1, such that the first eigenvalue, corresponding to the eigenvector $\ket{\psi}$, is -1. Finally, the desired Hamiltonian is constructed using eigendecomposition as:
\begin{equation}
    H_{\text{ctrl}}^{\psi} = Q\Lambda Q^{-1}.
\end{equation}
This Hamiltonian can be scaled by any real positive factor, while maintaining $\ket{\psi}$ as the ground state. 

\section*{Supplementary Note 2: Second-order Dyson expansion of the noise Operator ($V_O$)}
Given the total Hamiltonian of the system and bath in the form,
\begin{equation}
    H(t) = H_{\text{ctrl}}(t) + H_{\text{noise}}(t).
\end{equation}
Moving to the interaction picture with respect to the control Hamiltonian, we can express the total unitary as
\begin{align}
    U(t) &= \mathcal{T}_{+}\left(e^{-i\int_{0}^{t}H(s)ds}\right) := U_{\text{ctrl}}(t) U_I(t),
\end{align}
where the control unitary is given by
\begin{align}
    U_{\text{ctrl}}(t) &= \mathcal{T}_{+} \left(e^{-i\int_{0}^{t}H_{\text{ctrl}}(s)ds}\right),
\end{align}
and the interaction Hamiltonian in the interaction picture is given by,
\begin{align}    
    H_I(t) &= (U_{\text{ctrl}}^{\dagger}(t) H_{\text{noise}}(t) U_{\text{ctrl}}(t)).
\end{align}
Assume the joint initial state of system-bath is $\rho_S(0)\otimes\rho_B$. We can express the expectation of an observable $O$ at time $t=T$ as
\begin{align}
    \braket{O(T)} &= \braket{\text{Tr}_{SB}[U(T)(\rho_S(0)\otimes\rho_B)U^{\dagger}(T)O]}_c\\
  &=\braket{\text{Tr}_{SB}[U_{\text{ctrl}}(T)U_I(T)(\rho_S(0)\otimes\rho_B)U_I^{\dagger}(T)U_{\text{ctrl}}^{\dagger}(T)O]}_c\\
  &=\braket{\text{Tr}_{SB}[U_I^{\dagger}(T)U_{\text{ctrl}}^{\dagger}(T)OU_{\text{ctrl}}(T)U_I(T)U_{\text{ctrl}}^{\dagger}(T)U_{\text{ctrl}}(T)(\rho_S(0)\otimes\rho_B)U_{\text{ctrl}}^{\dagger}(T)U_{\text{ctrl}}(T)]}\\
 &:=\braket{\text{Tr}_{SB}[\tilde{U}_I^{\dagger}(T)O\tilde{U}_I(T)(U_{\text{ctrl}}(T)(\rho_S(0)\otimes\rho_B)U_{\text{ctrl}}^{\dagger}(T))OO^{-1}]}\\
 &=\braket{\text{Tr}_S[O^{-1}\text{Tr}_B[(\tilde{U}_I^{\dagger}(T)O\tilde{U}_I(T))\rho_B](U_{\text{ctrl}}(T)\rho_S(0) U_{\text{ctrl}}^{\dagger}(T))O]}_c\\
  &= \text{Tr}_S[\braket{O^{-1}\text{Tr}_B[(\tilde{U}_I^{\dagger}(T)O\tilde{U}_I(T))\rho_B]}_c U_{\text{ctrl}}(T)\rho_S(0) U_{\text{ctrl}}^{\dagger}(T)O]\\
  &:=\text{Tr}_S[V_O(T)U_{\text{ctrl}}(T)\rho_S(0)U_{\text{ctrl}}^{\dagger}(T)O],
\end{align}
In the second line, we use the definition of the interaction picture unitary. In the third line, we apply the cyclic property of the trace, and multiply twice by $U_{\text{ctrl}}^{\dagger}(T)U_{\text{ctrl}}(T)=\mathbb{I}_S$. In the fourth line, we introduce the modified interaction picture $\tilde{U}_I=U_{\text{ctrl}}(T)U_I(T)U_{\text{ctrl}}^{\dagger}(T)$ and multiply by $OO^{-1}=\mathbb{I}$ at the end. In the fifth line, we apply the cyclic property of trace and also take system operators out of the partial trace over the bath. In the sixth line, we take the system operators out of the classical averaging, since they are deterministic. 

Next, we perturbatively expand $V_O(T)$ by using the Dyson expansion of interaction unitary $U_I(T)$, up to the second-order in $H_I(t)$, as follows:
\begin{align}
    V_O(T)&=O^{-1}\braket{\text{Tr}_B[(\tilde{U}_I^{\dagger}O\tilde{U}_I)(\mathbb{I}\otimes\rho_B)]}_c\\
    &=O^{-1}\text{Tr}_B[\braket{U_{\text{ctrl}}(T)U_I^{\dagger}(T)U_{\text{ctrl}}^{\dagger}(T)OU_{\text{ctrl}}(T)U_I(T)U_{\text{ctrl}}^{\dagger}(T)}_c \rho_B]\\
    &\approx O^{-1}\text{Tr}_B[\braket{U_{\text{ctrl}}(T)(\mathbb{I}+D_1^{\dagger}+D_2^{\dagger})U_{\text{ctrl}}^{\dagger}(T)OU_{\text{ctrl}}(T)(\mathbb{I} + D_1 + D_2)U_{\text{ctrl}}^{\dagger}(T)}_c \rho_B]\\
    &= O^{-1}\text{Tr}_B[\braket{(\mathbb{I}+\tilde{D}_1^{\dagger}+\tilde{D}_2^{\dagger})O(\mathbb{I} + \tilde{D}_1 + \tilde{D}_2)}_c \rho_B]\\
    &\approx \mathbb{I}+\text{Tr}_B[\braket{O^{-1}\tilde{D}_1^{\dagger}O+\tilde{D}_1+O^{-1}\tilde{D}_1^{\dagger}O\tilde{D}_1+O^{-1}\tilde{D}_2^{\dagger}O+\tilde{D}_2}_c\rho_B],
\end{align}
In the third line, we expand the interaction unitary into Dyson terms up to second-order. In the fourth line, we define $\tilde{D}_i=U_{\text{ctrl}}(T)D_i(T)U_{\text{ctrl}}^{\dagger}(T)$, where 
\begin{align}
    {D}_1 &= -i \int_{0}^{T}ds \text{ } {H}_I(s)\\
    {D}_1^{\dagger} &= i \int_{0}^{T}ds \text{ } {H}_I(s)\\
    {D}_2 &= -\int_{0}^{T}ds\int_0^s ds' {H}_I(s){H}_I(s')\\
    {D}_2^{\dagger}&= -\int_0^Tds \int_0^s ds' {H}_I(s'){H}_I(s).
\end{align}
In the fifth line, we only consider the terms which are up to second-order in $H_I(t)$. If the system is subject to classical noise only, then $H_{\text{noise}}(t)$ is a system-only operator, $\rho_B = \mathbb{I}$, and thus the expression of the $V_O(T)$ operator is reduced to
\begin{equation}\label{vo:expr}
    V_O(T) \approx \mathbb{I}+\braket{O^{-1}\tilde{D}_1^{\dagger}O+\tilde{D}_1+O^{-1}\tilde{D}_1^{\dagger}O\tilde{D}_1+O^{-1}\tilde{D}_2^{\dagger}O+\tilde{D}_2}_c.
\end{equation}

\section*{Supplementary note 3: showing that $C_3>0$}
Here, we show that
\begin{align}\label{ineq:exact}
    C_3(s)=1-e_1^2(s)-e_2^2(s)>0\ \forall s\in [0,1],
\end{align}
where $e_{1}(s), e_{2}(s)$ are given by
\begin{align}
    e_{1}(s) &= r_1(t_i) + \left(r_1(t_f) - r_1(t_i) \right)(3s^2-2s^3) \\
    e_{2}(s) &= r_2(t_i) + \left(r_2(t_f) - r_2(t_i) \right)(3s^2-2s^3),
\end{align}
and $r_k(t)=h_k(t)/R(t)$ and $R(t) = \sqrt{h_1^2(t) + h_2^2(t) + h_3^2(t)}$. The two functions $e_1(s)$ and $e_2(s)$ are monotonic functions. If $r_k(t_f)-r_k(t_i)=0$, $e_k(s)$ is constant, which is monotonic trivially. Otherwise, we compute the derivative as 
\begin{align}
    \dot{e}_{1}(s) &= 6\left(r_1(t_f) - r_1(t_i) \right)s(1-s) \\
    \dot{e}_{2}(s) &= 6\left(r_2(t_f) - r_2(t_i) \right)s(1-s).
\end{align}
Each derivatives vanishes only at $s=0,1$, and will have the same sign in between. Thus $e_1(s),e_2(s)$ are monotonic in $s\in [0,1]$. As a result, going from $s=0$ to $s=1$, $e_k(s)$ monotonically goes from $e_k(0)$ to $e_k(1)$. Now, comparing $e_1(s)$ and $e_2(s)$, we find that
\begin{align}
    3s^2-2s^3&=\frac{e_1(s)-r_1(s)}{r_1(t_f) - r_1(t_i)}=\frac{e_2(s)-r_2(s)}{r_2(t_f) - r_2(t_i)},
\end{align}
or, 
\begin{align}
    e_2(s)&=\frac{r_2(t_f) - r_2(t_i)}{r_1(t_f) - r_1(t_i)}(e_1(s)-r_1(t_i)) + r_2(t_i)\\
    &:=m(e_1(s)-r_1(t_i)) + r_2(t_i) \\
    &=me_1(s) + (r_2(t_i) - m r_1(t_i)).
\end{align}
Now, we have an implicit linear relation between $e_2(s)$ in terms of $e_1(s)$. This allows us to write $C_3(s)$ in terms of $e_1(s)$, by substituting in Equation \ref{ineq:exact}:
\begin{align}
    C_3(s)&=1-e_1^2(s)-\left(me_1(s) + (r_2(t_i) - m r_1(t_i)) \right)^2\\
    &=1-e_1^2(s)-m^2e_1^2(s) -2me_1(s)(r_2(t_i) - m r_1(t_i)) - (r_2(t_i) - m r_1(t_i))^2 \\
    &=-e_1^2(s)(1+m^2) - e_1(s)\left(2m(r_2(t_i) - m r_1(t_i)) \right) - (r_2(t_i) - m r_1(t_i))^2.
\end{align}
This shows that $C_3(s)$ is a quadratic function of $e_1(s)$. Since the coefficient of quadratic term is strictly negative, the quadratic is concave downwards. Looking at the boundaries, we have that
\begin{align}
    C_3(0) &= 1 - e_1^2(0) - e_2^2(0)\\
    &=1 - \frac{h_1^2(t_i)+h_2^2(t_i)}{h_1^2(t_i)+h_2^2(t_i)+h_3^2(t_i)} \\
    &= \frac{h_3^2(t_i)}{h_1^2(t_i)+h_2^2(t_i)+h_3^2(t_i)}>0,
\end{align}
where we have to choose $h_3(t_i)\neq 0$. Similarly we also obtain that $C_3(1) > 0$, if $h_3(t_f)\neq 0$. Because of the monotonicity of $e_1(s)$, we see that going from $s=0$ to $s=1$, $C_3(s)$ goes from a positive value at $s=0$ to another positive value at $s=1$. Finally, the concavity of $C_3(s)$, implies that $C_3(s)>0,\ \forall s \in [0,1]$. \qed 

\section*{Supplementary note 4: Optimization algorithm}
The algorithm starts by computing the cost function for an initial set of points $v_{k,j}/v_{\max}$, and choosing the point with the minimum cost to be the starting point. Then an iterative procedure is performed where either this point is retained, or replace by a new point. The choice between these two alternatives is based on constructing a hyperrectangle centered around the current point with the length of the $k,j$ side set to be $2\times\min\{(1-v_{k,j}/v_{\max}),(1+v_{k,j}/v_{\max})\}$. We then choose 100 uniformly-distributed random points within this hyperrectangle and evaluate  the cost function for each point.  Finally, we choose the minimum over all points, and compare it to the current center point.  If the new cost is less than the current one, the new point is considered as new center and a new hyperrectangle is constructed around it for the next iteration. On the other hand, if the new cost is higher, we retain the center and we proceed to the next iteration, with hyperrectangle sides shrunk by 5\%. We stop the algorithm when reaching a set maximum number of iterations, and the optimal point is the center of the last hyperrectangle. 

\section*{Supplementary Note 5: Second-order analytic expression of the noise operator for multi-axis RTN noise}
The RTN noise process is a binary process switching between two states at $t\in [0, T]$ with probability $\gamma$. The RTN process can thus be defined as
\begin{equation}
\beta(t) = \beta(0)(-1)^{n(0,t)},
\end{equation}
where $n(0,t)$ represents the number of switches during the time interval $[0,t]$, following a Poisson distribution with $\langle n(0,t)\rangle = \gamma t$. The initial value $\beta(0)=\pm 1$ is chosen randomly by flipping an unbiased coin. All odd-ordered moments of $\beta(t)$ vanish, and the second order moment is given by $\braket{\beta(t_1)\beta(t_2)}=e^{-2\gamma(t_1-t_2)}$.

Now, following Equation (62) in the main text, we have
\begin{equation}
    H_{\text{noise}}(t) = g_x\beta_x(t)\sigma_x + g_z\beta_z(t)\sigma_z.
\end{equation}
$\beta_x(t)$ and $\beta_z(t)$ are two independent RTN noise processes. Thus, the correlation is given by $\braket{\beta_i(t_1)\beta_j(t_2)}=e^{-2\gamma(t_1-t_2)}\delta_{ij}$. The interaction Hamiltonian in the interaction picture is given by,
\begin{align}    
    H_I(t) &= g_x \beta_x(t)(U_{\text{ctrl}}^{\dagger}(t) \sigma_x U_{\text{ctrl}}(t))+g_z \beta_z(t)(U_{\text{ctrl}}^{\dagger}(t) \sigma_z U_{\text{ctrl}}(t))\\
    &= g_x\sum_{a\in \{0,x,y,z\}} y_{ax}(t)\sigma_a\beta_x(t) + g_z\sum_{a\in \{0,x,y,z\}} y_{az}(t)\sigma_a\beta_z(t)\\
    y_{ax}(t)  &:= \text{Tr}[U_{\text{ctrl}}^{\dagger}(t)\sigma_x U_{\text{ctrl}}(t) \sigma_a]/2\\
    y_{az}(t)  &:= \text{Tr}[U_{\text{ctrl}}^{\dagger}(t)\sigma_z U_{\text{ctrl}}(t) \sigma_a]/2,
\end{align}
where $\sigma_0=\mathbb{I}_S$. Here we use the fact that Pauli operators form an orthonormal set of basis for any $2\times 2$ operator. We can then calculate the terms in Equation \ref{vo:expr} as follows.
\begin{align}
     \braket{O^{-1}\tilde{D}_1^{\dagger}O} &= O^{-1}\braket{\tilde{D}_1^{\dagger}}O\\
    &= -i \int_0^T \left( \sum_a (g_x y_{ax}(s) \braket{\beta_x(s)} + g_z y_{az}(s) \braket{\beta_z(s)}) O^{-1} \tilde{\sigma}_a O\right) ds\\
    &=0,
\end{align}
where $\tilde{\sigma_a}=U_{\text{ctrl}}(T)\sigma_a U_{\text{ctrl}}^\dagger(T)$ and the noise has zero mean. Similarly, we have $\braket{\tilde{D}_1} =0 $. Next, we have
\begin{align}
    O^{-1}\braket{\tilde{D}_1^{\dagger}O\tilde{D}_1} &= \sum_{a,a'}O^{-1}\tilde{\sigma}_aO\tilde{\sigma}_{a'}\int_0^T \int_0^{T}\bigg( y_{ax}(r)y_{a'x}(r') \braket{g_x\beta_x(r)g_x\beta_x(r')} \nonumber\\ \nonumber&+y_{ax}(r)y_{a'z}(r') \braket{g_x\beta_x(r)g_z\beta_z(r')}\\ \nonumber&+ y_{az}(r)y_{a'x}(r') \braket{g_z\beta_z(r)g_x\beta_x(r')} \\ &+y_{az}(r)y_{a'z}(r') \braket{g_z\beta_z(r)g_z\beta_z(r')} \bigg)dr'dr\\
    &= g_x^2\sum_{a,a'} I_{ax,a'x}O^{-1}\tilde{\sigma}_aO\tilde{\sigma}_{a'} + g_z^2\sum_{a,a'} I_{az,a'z}O^{-1}\tilde{\sigma}_aO\tilde{\sigma}_{a'},
\end{align}
where the $I_{ax,a'x}$ and $I_{az,a'z}$ can be computed numerically given the prior knowledge of the second-order correlation function of the noise $\braket{\beta_x(t_1)\beta_x(t_2)}$ and $\braket{\beta_z(t_1)\beta_z(t_2)}$, respectively. The two remaining terms are
\begin{align}
    \braket{O^{-1}\tilde{D}_2^{\dagger}O} &= O^{-1}\braket{\tilde{D}_2^{\dagger}}O\\
    &=-\sum_{a,a'}O^{-1}\tilde{\sigma}_a\tilde{\sigma}_{a'}O\int_0^T \int_0^{r}\bigg(y_{ax}(r)y_{a'x}(r') \braket{g_x\beta_x(r)g_x\beta_x(r')} \nonumber\\ \nonumber&+y_{ax}(r)y_{a'z}(r') \braket{g_x\beta_x(r)g_z\beta_z(r')}\\ \nonumber&+ y_{az}(r)y_{a'x}(r') \braket{g_z\beta_z(r)g_x\beta_x(r')} \\ &+y_{az}(r)y_{a'z}(r') \braket{g_z\beta_z(r)g_z\beta_z(r')}\bigg) dr'dr\\
    &:= -g_x^2\sum_{a,a'} I_{ax,a'x}^{>} O^{-1}\tilde{\sigma}_a\tilde{\sigma}_{a'}O - g_z^2\sum_{a,a'} I_{az,a'z}^{>} O^{-1}\tilde{\sigma}_a\tilde{\sigma}_{a'}O\\
    \braket{\tilde{D}_2} &= -\sum_{a,a'}\tilde{\sigma}_a\tilde{\sigma}_{a'}\int_0^T \int_0^{r} \bigg(y_{ax}(r')y_{a'x}(r) \braket{g_x\beta_x(r')g_x\beta_x(r)} \nonumber\\ \nonumber&+y_{ax}(r')y_{a'z}(r) \braket{g_x\beta_x(r')g_z\beta_z(r)}\\ \nonumber&+ y_{az}(r')y_{a'x}(r) \braket{g_z\beta_z(r')g_x\beta_x(r)} \\ &+y_{az}(r')y_{a'z}(r) \braket{g_z\beta_z(r')g_z\beta_z(r)}\bigg)dr'dr \\
    &:= -g_x^2\sum_{a,a'} I_{ax,a'x}^{<}\tilde{\sigma}_a\tilde{\sigma}_{a'} - g_z^2\sum_{a,a'} I_{az,a'z}^{<}\tilde{\sigma}_a\tilde{\sigma}_{a'},
\end{align}
where $I_{a,a'}^{>}$ and $I_{a,a'}^{<}$ can similarly be computed numerically. This gives an analytic expression of the noise operator for independent multi-axis RTN noise up to second-order.

\section*{Supplementary note 6: Graybox architecture}
In this paper, we use the following graybox architecture. The input to the model is the pulse parameters $\vec{\Theta}$ and the output is the expectation values of observables ($\braket{O(T)}$). In general, the noise operator $V_O(T)$ can be eigendecomposed as 
\begin{equation}\label{eq:gb1}
    V_O(T) = O^{-1}QDQ^\dagger,
\end{equation}
where the matrices $Q$ and $D$ can be parameterized as \cite{W3}
\begin{equation}\label{eq:gb2}
    Q = \begin{pmatrix}
        e^{i\psi}&0\\
        0&e^{-i\psi}
    \end{pmatrix}\begin{pmatrix}
        \cos\phi&\sin\phi\\
        -\sin\phi&\cos\phi
    \end{pmatrix}\begin{pmatrix}
        e^{i\Delta}&0\\
        0&e^{-i\Delta}
    \end{pmatrix}
\end{equation}
\begin{equation}\label{eq:gb3}
    D = \begin{pmatrix}
        \mu_1&0\\
        0&\mu_2
    \end{pmatrix}.
\end{equation}
The parameters $\psi, \phi, \Delta, \mu_1$ and $\mu_2$ can be mapped from $\vec{\Theta}$. In general, finding an exact analytic expression of this mapping might not be possible. Thus, we use a blackbox structure to approximately model this map. The blackbox consists of an initial GRU with 60 nodes, connected to the input pulse parameters. This is followed by a GRU for each Pauli operator consisting of 60 nodes. Finally, each of these GRUs are connected to a neural layer with five output nodes. The first three nodes have a linear activation function with their output representing $\psi, \phi$ and $\Delta$ which can be used to construct $Q$, defined earlier. The other two nodes have hyperbolic tangent (tanh) activation function and correspond to $\mu_1$ and $\mu_2$ parameters. These activations are chosen to satisfy the constraints on the parameters, as explained in \cite{W1}. The whitebox part starts with a layer that constructs the noise operators from the parameters obtained from the neural layers. The next layer computes the expectations of observables using the expression:
\begin{equation}
    \braket{O(T)} = \text{Tr}[V_O\ket{\psi_T}\bra{\psi_T}O].
\end{equation}
\begin{figure}
    \centering
    \includegraphics[width=0.65\linewidth]{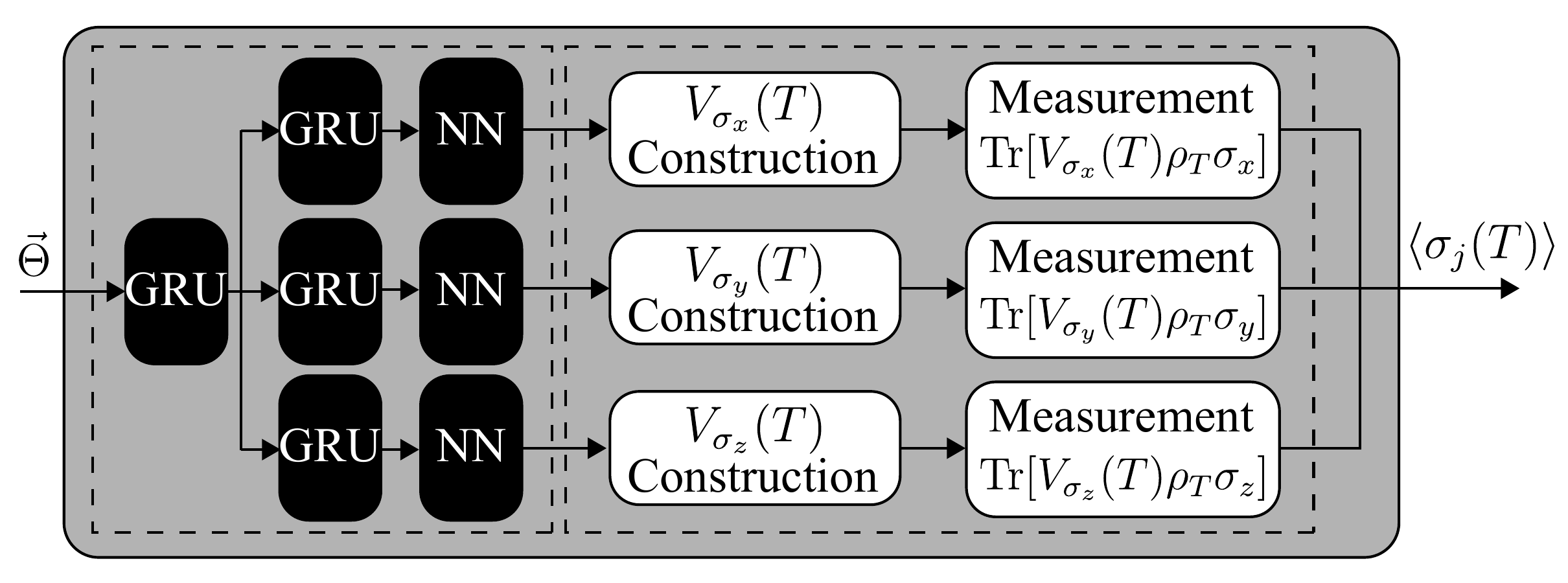}
    \caption{\textbf{The proposed graybox model.} The architecture consists of blackbox and whitebox structures. The blackbox consists of GRUs and neural networks, which outputs the matrix parameters of $V_{\sigma_x}(T),V_{\sigma_y}(T)$, and $V_{\sigma_z}(T)$. The first whitebox layer constructs the noise operators from the outputs of the blackbox. The measurement layers compute the expectations of each Pauli operator, with target state $\rho_T=\ket{\psi_T}\!\!\bra{\psi_T}$.}
    \label{fig:arch}
\end{figure}

\bibliographystyle{unsrt}
\bibliography{bib}

\newpage

\begin{table}[t]
\caption{\textbf{The bound $\mathbf{v_{\textbf{max}}/c}$ on the free variables in each subtrajectory for the different target states for the qubit system.} The reference axis for subtrajectory 1 is $Z$ for all targets; for subtrajectory 2 it is $X$ for targets (iii), (iv) and (vi), while $Y$ for target (v); and for subtrajectory 3 it is $Z$ for target (vi).}
\label{table:vmax2}
\centering
\begin{tabular}{|c|c|c|c|}
\hline
\textbf{Target} & \textbf{Subtrajectory 1} & \textbf{Subtrajectory 2} & \textbf{Subtrajectory 3} \\
\hline
 (i)  & 0.1655 & - & -\\
(ii)  & 0.6352 & - & -\\
(iii) & 0.1564 & 0.3518 & -\\
(iv)  & 0.1951 & 0.2761 & -\\
(v)   & 0.1832 & 0.4800 & -\\
(vi)  & 0.2985 & 0.3481 & 0.5739 \\
\hline
\end{tabular}
\end{table}

\section*{Supplementary Figures}

\begin{figure*}[h]
    \centering
    \includegraphics[width=\linewidth]{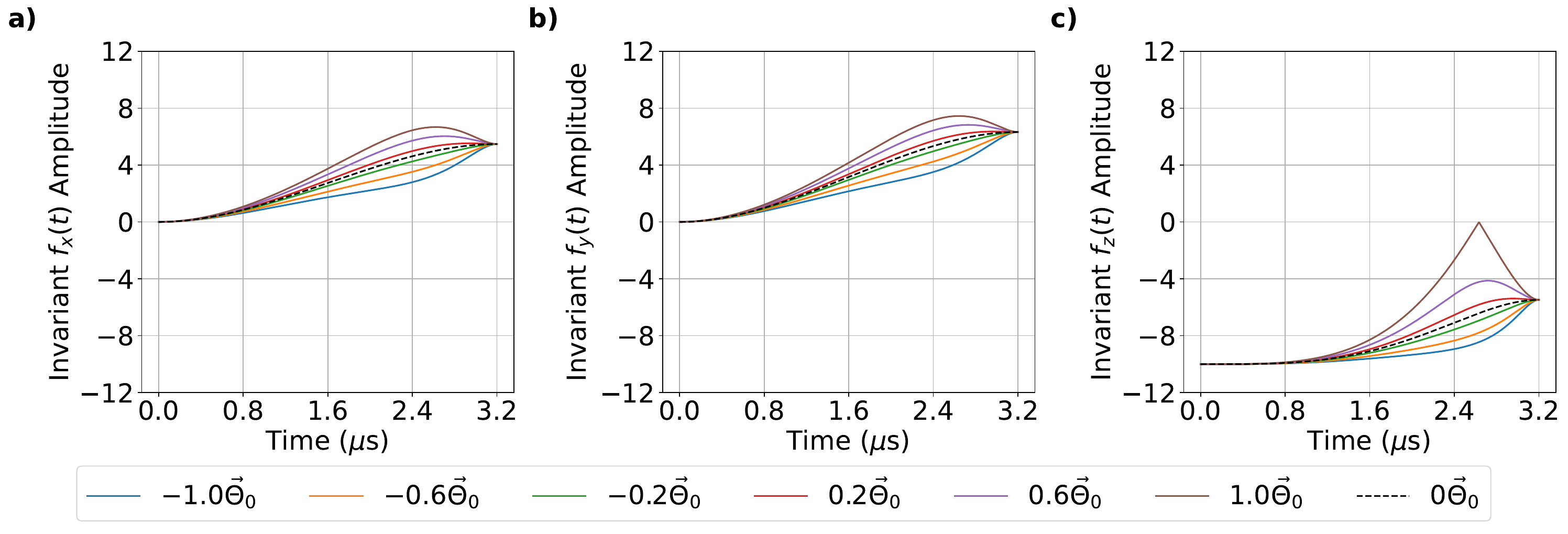}
    \caption{\textbf{Family of invariants for target (i) for the qubit system.} This case requires 1 subtrajectory. Different examples from the invariant family are plotted as function of time. The parameters are chosen to be of linear scaling of $\vec{\Theta}_0$ that corresponds to the extreme case where $f_3(t)\to 0$ at exactly 1 point. The dotted curve represents degree 3 solution, i.e. the minimal polynomial that satisfies the boundary conditions. }
    \label{fig:samefam}
\end{figure*}

\begin{figure*}
    \centering
    \includegraphics[width=\linewidth]{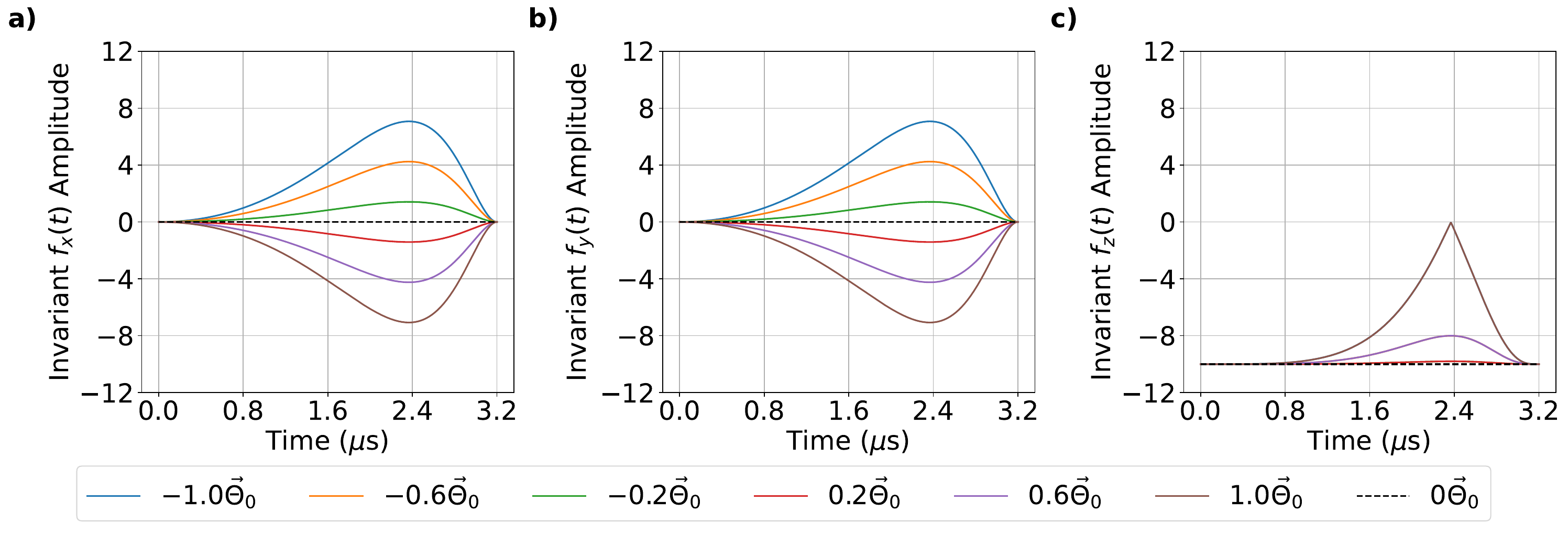}
    \caption{\textbf{Family of invariants for target (ii) for the qubit system.} This case requires 1 subtrajectory. Different examples from the invariant family are plotted as function of time. The parameters are chosen to be of linear scaling of $\vec{\Theta}_0$ that corresponds to the extreme case where $f_3(t)\to 0$ at exactly 1 point. The dotted curve represents degree 3 solution, i.e. the minimal polynomial that satisfies the boundary conditions. }
    \label{fig:memfam}
\end{figure*}

\begin{figure*}
    \centering
    \includegraphics[width=\linewidth]{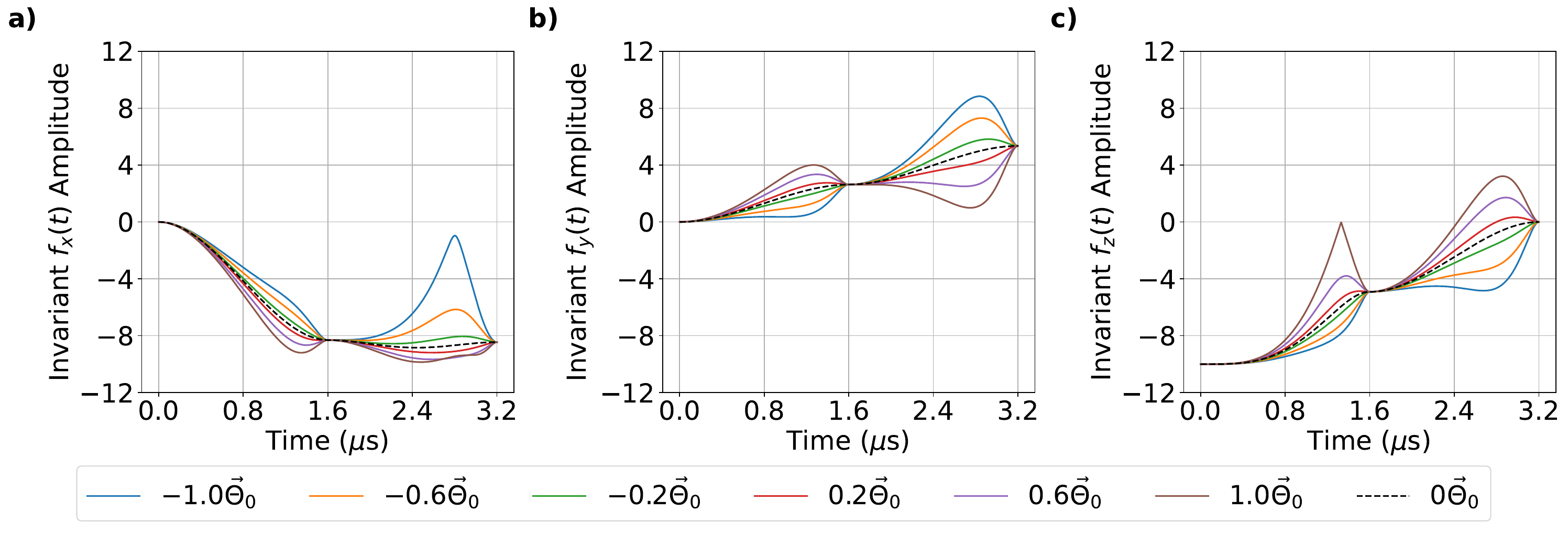}
    \caption{\textbf{Family of invariants for target (iii) for the qubit system.} This case requires 2 subtrajectories. Different examples from the invariant family are plotted as function of time. The parameters are chosen to be of linear scaling of $\vec{\Theta}_0$ that corresponds to the extreme case where $f_3(t)\to 0$ at exactly 1 point. The dotted curve represents degree 3 solution, i.e. the minimal polynomial that satisfies the boundary conditions. }
    \label{fig:equfam}
\end{figure*}

\begin{figure*}
    \centering
    \includegraphics[width=\linewidth]{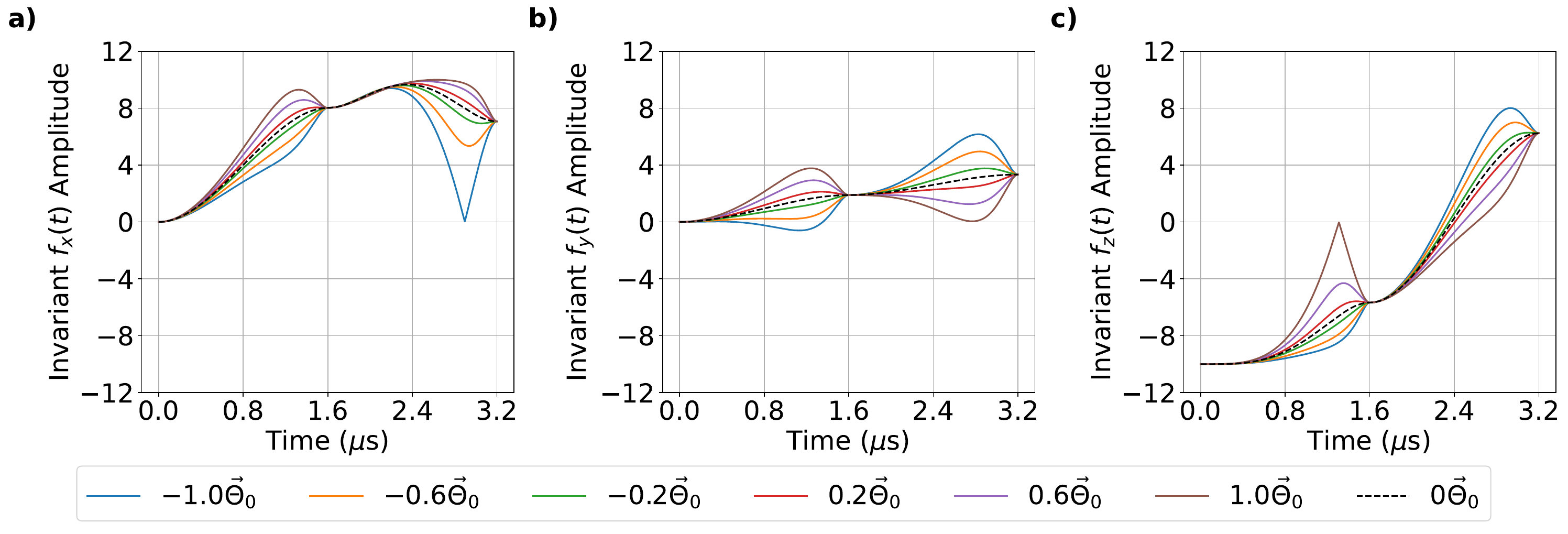}
    \caption{\textbf{Family of invariants for target (iv) for the qubit system.} This case requires 2 subtrajectories. Different examples from the invariant family are plotted as function of time. The parameters are chosen to be of linear scaling of $\vec{\Theta}_0$ that corresponds to the extreme case where $f_3(t)\to 0$ at exactly 1 point. The dotted curve represents degree 3 solution, i.e. the minimal polynomial that satisfies the boundary conditions. }
    \label{fig:travfam}
\end{figure*}

\begin{figure*}
    \centering
    \includegraphics[width=\linewidth]{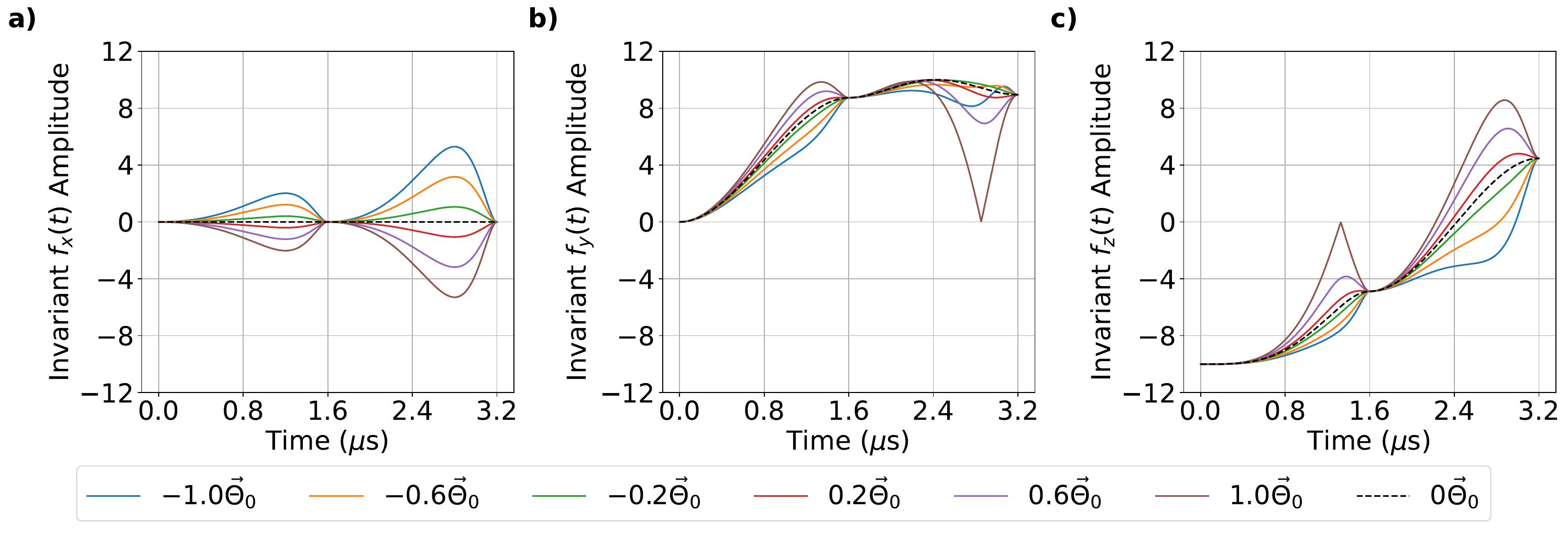}
    \caption{\textbf{Family of invariants for target (v) for the qubit system.} This case requires 2 subtrajectories. Different examples from the invariant family are plotted as function of time. The parameters are chosen to be of linear scaling of $\vec{\Theta}_0$ that corresponds to the extreme case where $f_3(t)\to 0$ at exactly 1 point. The dotted curve represents degree 3 solution, i.e. the minimal polynomial that satisfies the boundary conditions. }
    \label{fig:trav_yfam}
\end{figure*}

\begin{figure*}
    \centering
    \includegraphics[width=\linewidth]{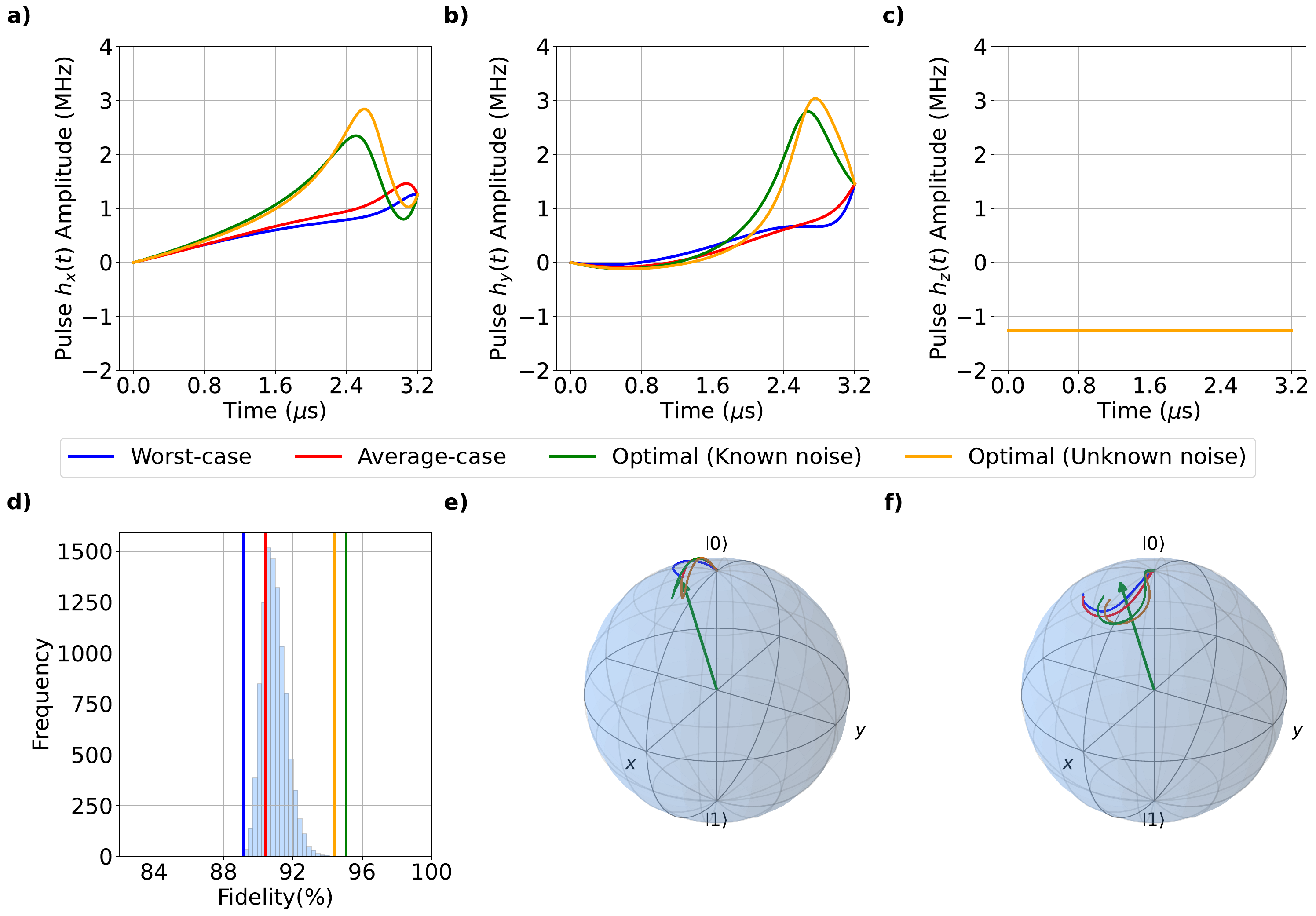}
    \caption{\textbf{Optimization results for target (i) for the qubit system.} We compare the optimization result (known and unknown noise) with the worst-case and average-case pulse in from the dataset. The optimal pulse in known noise setting is obtained using a whitebox approach, while graybox approach is used to find optimal pulse for unknown noise setting. a), b) and c) show the pulse waveforms as function of time along x, y and z-axis, respectively. d) shows the fidelity of these pulses, in relation to the histogram of fidelities computed over the whole dataset. e) shows the state trajectory generated by each of these pulses on the Bloch sphere for a closed system, while f) shows the open system trajectory.}
    \label{fig:sameres}
\end{figure*}

\begin{figure*}
    \centering
    \includegraphics[width=\linewidth]{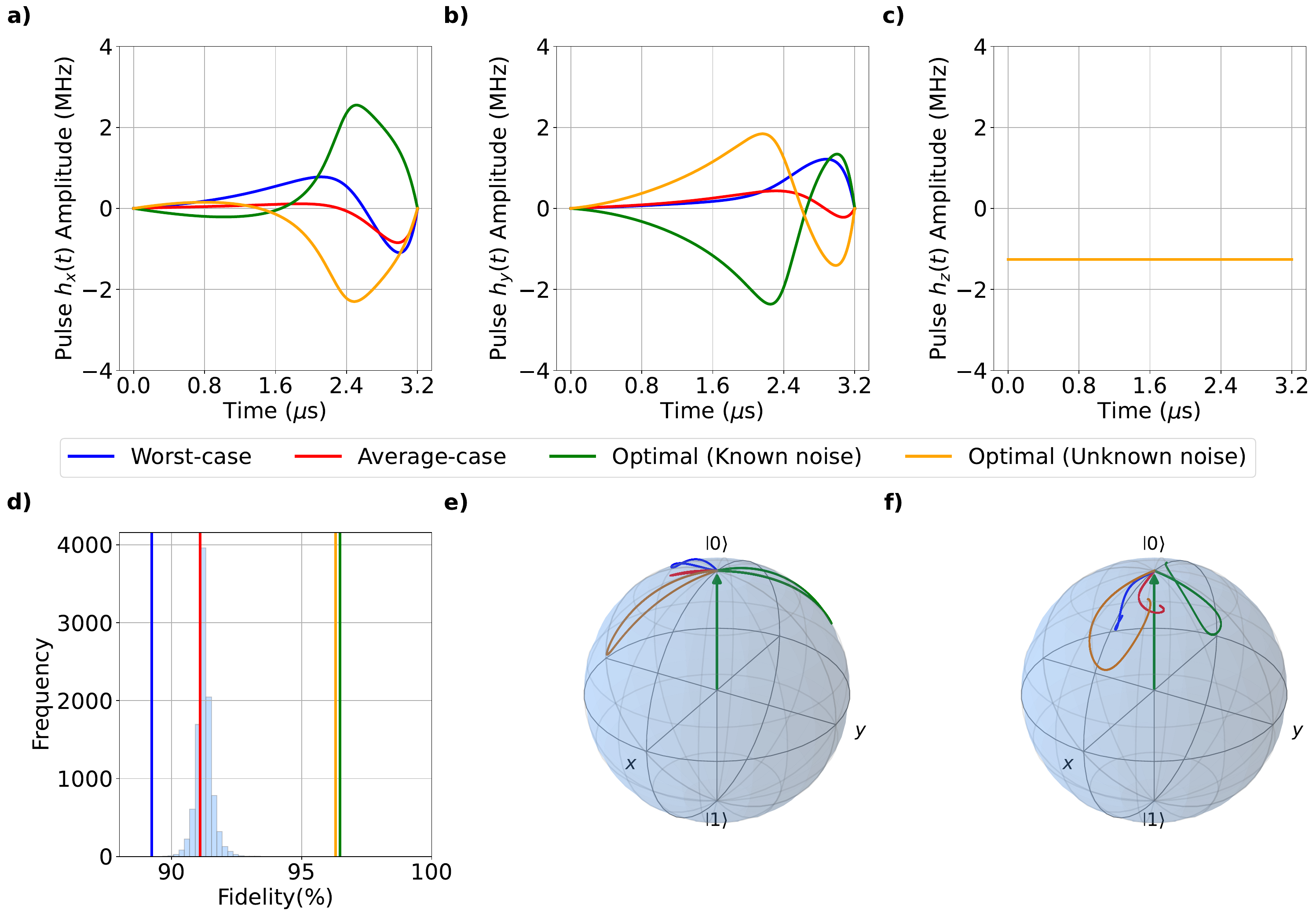}
    \caption{\textbf{Optimization results for target (ii) for the qubit system.} We compare the optimization result (known and unknown noise) with the worst-case and average-case pulse in from the dataset. The optimal pulse in known noise setting is obtained using a whitebox approach, while graybox approach is used to find optimal pulse for unknown noise setting. a), b) and c) show the pulse waveforms as function of time along x, y and z-axis, respectively. d) shows the fidelity of these pulses, in relation to the histogram of fidelities computed over the whole dataset. e) shows the state trajectory generated by each of these pulses on the Bloch sphere for a closed system, while f) shows the open system trajectory.}
    \label{fig:memres}
\end{figure*}

\begin{figure*}
    \centering
    \includegraphics[width=\linewidth]{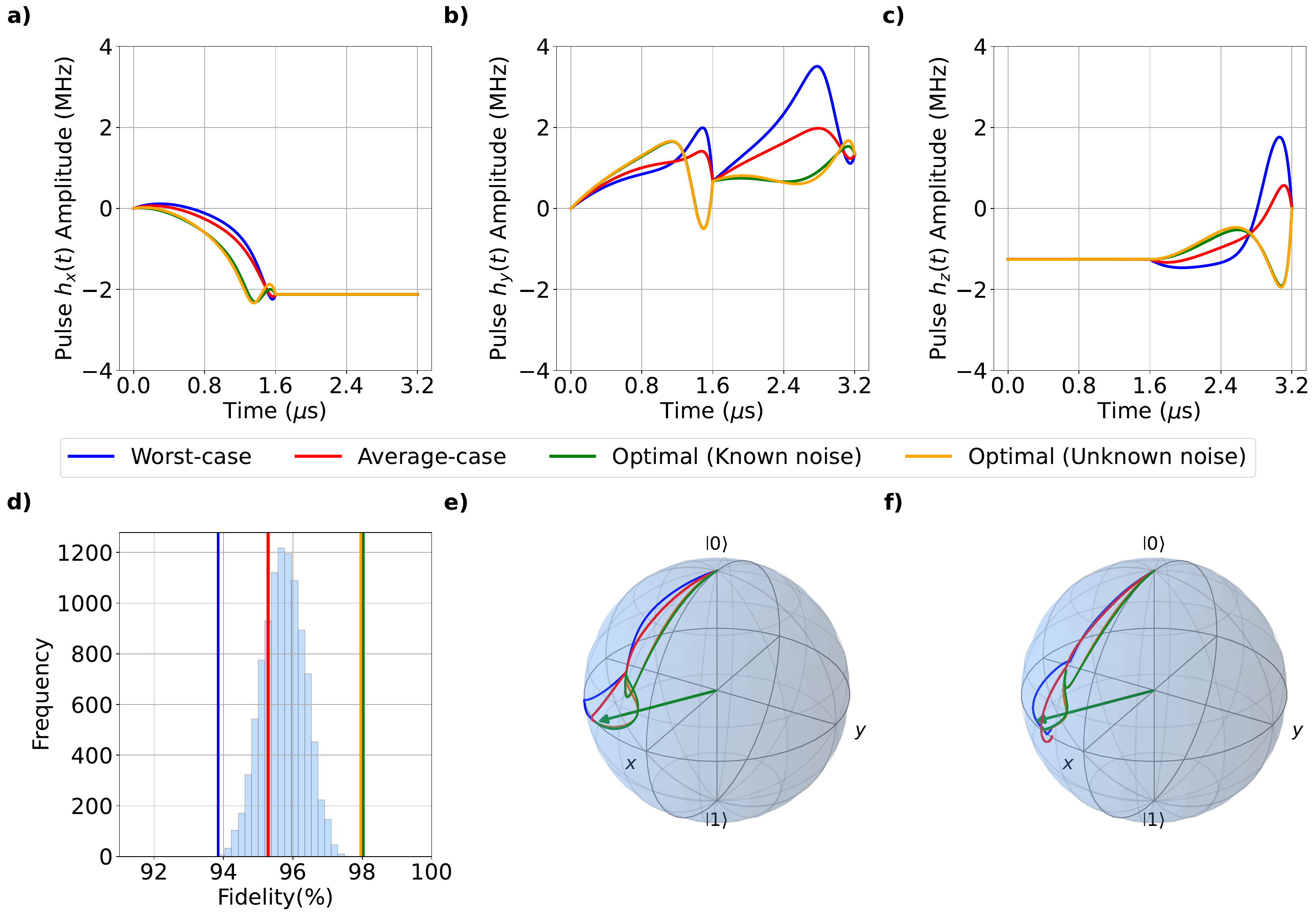}
    \caption{\textbf{Optimization results for target (iii) for the qubit system.} We compare the optimization result (known and unknown noise) with the worst-case and average-case pulse in from the dataset. The optimal pulse in known noise setting is obtained using a whitebox approach, while graybox approach is used to find optimal pulse for unknown noise setting. a), b) and c) show the pulse waveforms as function of time along x, y and z-axis, respectively. d) shows the fidelity of these pulses, in relation to the histogram of fidelities computed over the whole dataset. e) shows the state trajectory generated by each of these pulses on the Bloch sphere for a closed system, while f) shows the open system trajectory.}
    \label{fig:equres}
\end{figure*}

\begin{figure*}
    \centering
    \includegraphics[width=\linewidth]{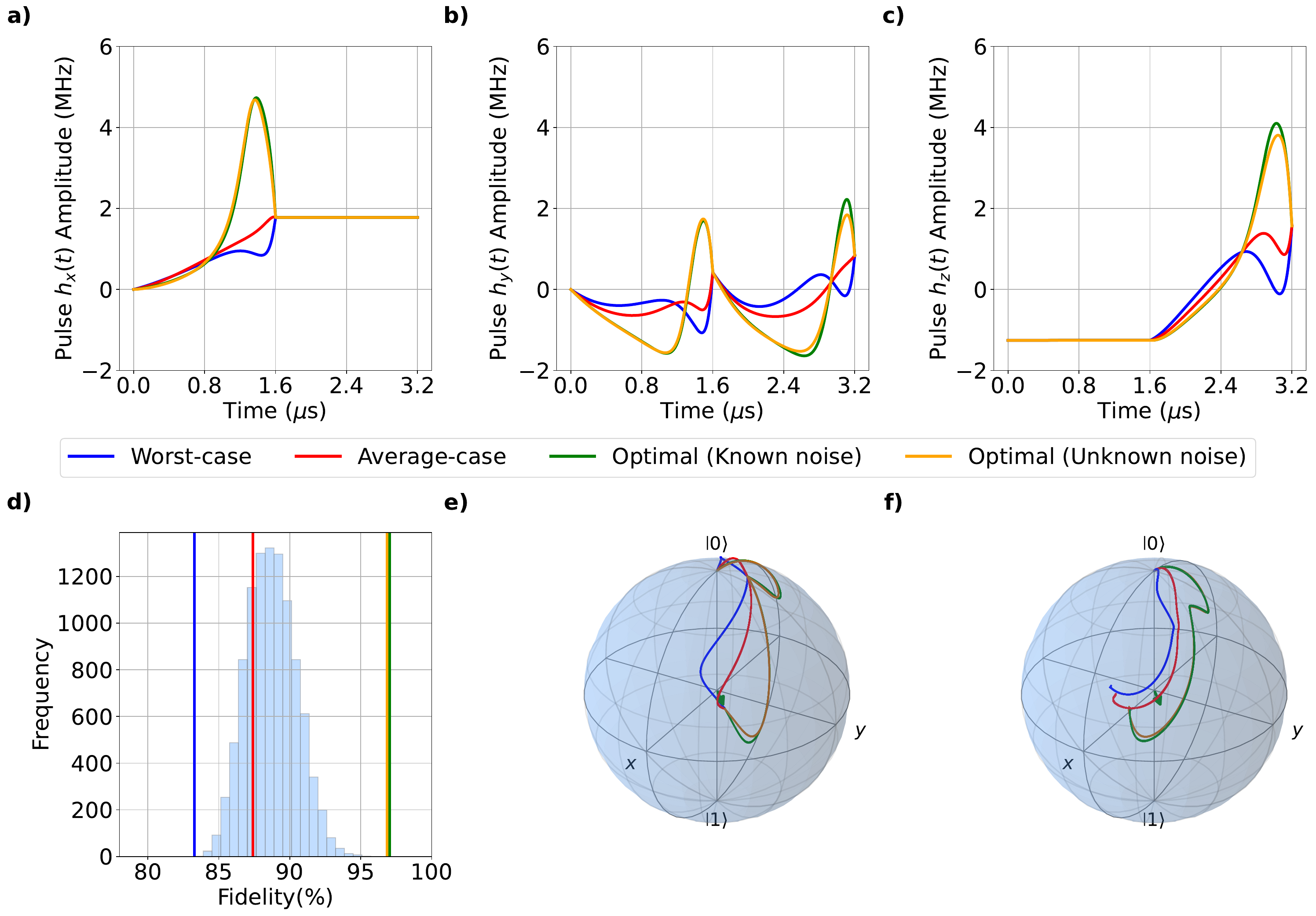}
    \caption{\textbf{Optimization results for target (iv) for the qubit system.} We compare the optimization result (known and unknown noise) with the worst-case and average-case pulse in from the dataset. The optimal pulse in known noise setting is obtained using a whitebox approach, while graybox approach is used to find optimal pulse for unknown noise setting. a), b) and c) show the pulse waveforms as function of time along x, y and z-axis, respectively. d) shows the fidelity of these pulses, in relation to the histogram of fidelities computed over the whole dataset. e) shows the state trajectory generated by each of these pulses on the Bloch sphere for a closed system, while f) shows the open system trajectory.}
    \label{fig:travres}
\end{figure*}

\begin{figure*}
    \centering
    \includegraphics[width=\linewidth]{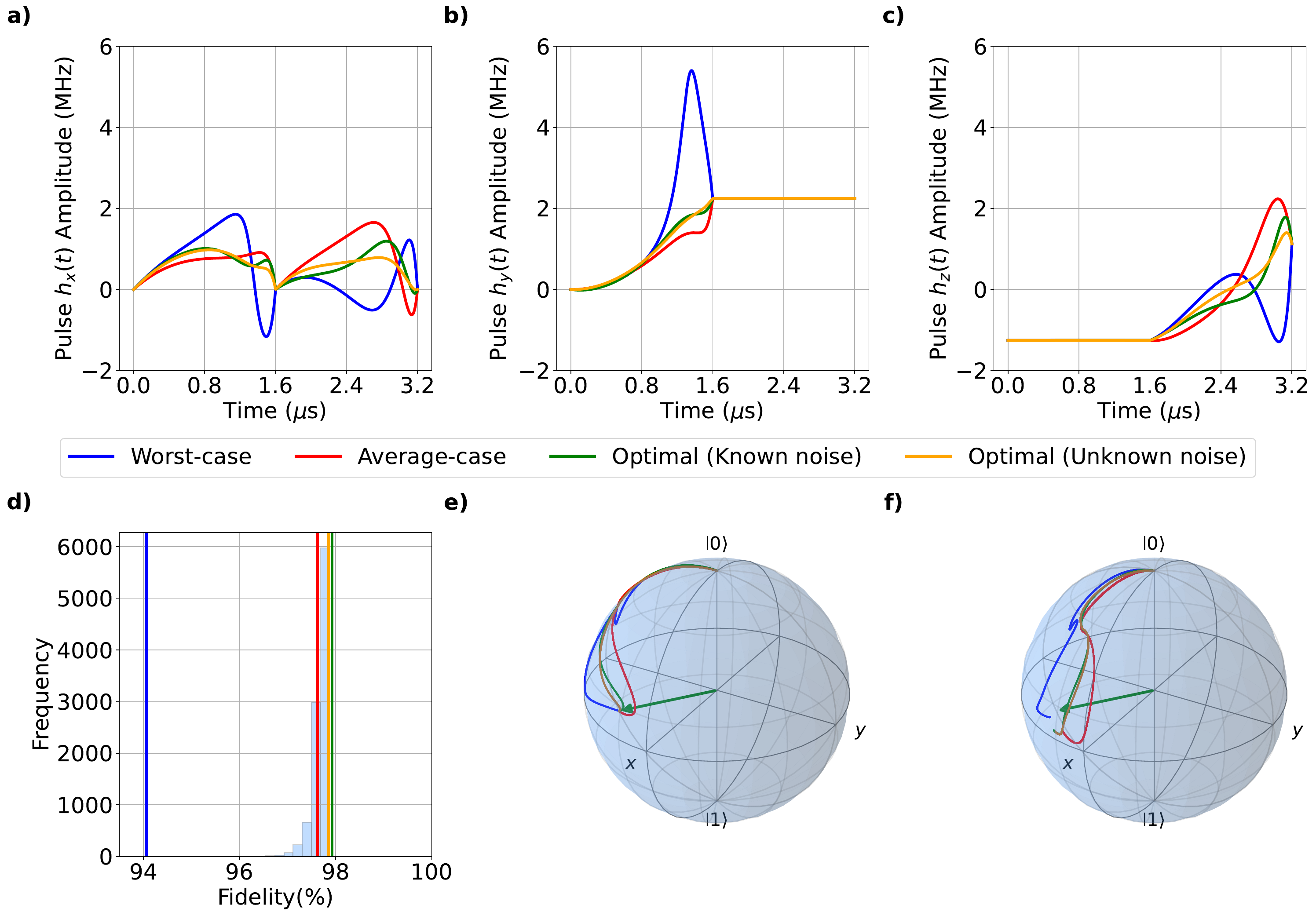}
    \caption{\textbf{Optimization results for target (v) for the qubit system.} We compare the optimization result (known and unknown noise) with the worst-case and average-case pulse in from the dataset. The optimal pulse in known noise setting is obtained using a whitebox approach, while graybox approach is used to find optimal pulse for unknown noise setting. a), b) and c) show the pulse waveforms as function of time along x, y and z-axis, respectively. d) shows the fidelity of these pulses, in relation to the histogram of fidelities computed over the whole dataset. e) shows the state trajectory generated by each of these pulses on the Bloch sphere for a closed system, while f) shows the open system trajectory.}
    \label{fig:trav_yres}
\end{figure*}

\begin{figure*}
    \centering
    \includegraphics[width=0.95\linewidth]{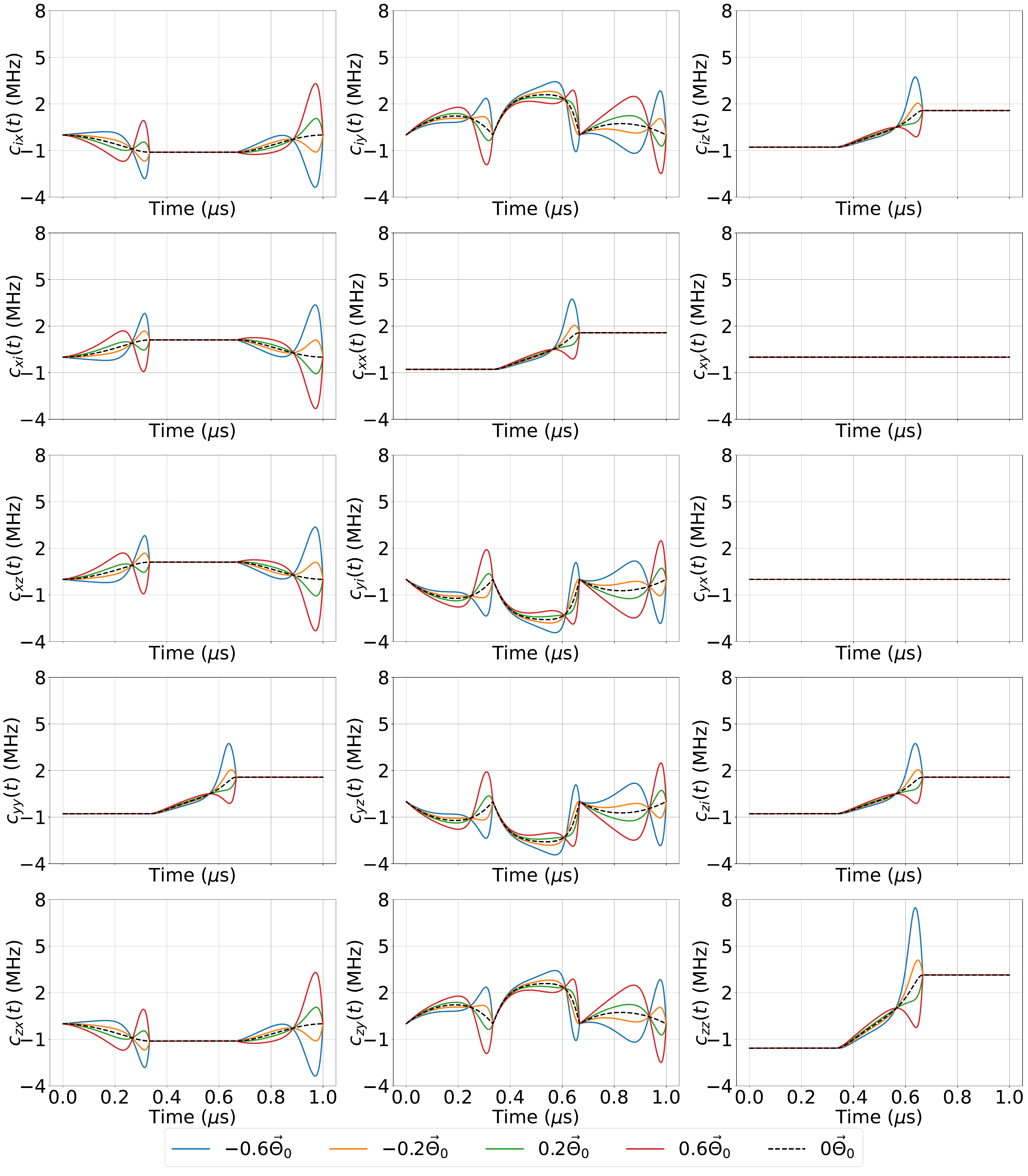}
    \caption{\textbf{The two-qubit system pulse waveforms for different members of the invariant family.} The pulse waveforms $c_{ij}(t)$ corresponding to the two-qubit Pauli basis $\sigma_i \otimes \sigma_j$ for ($i,j = 0,x,y,z$) are shown for different members of the pulse family (we skip $C_{00}(t)$ as it corresponds to the identity operator, which only results in a global phase shift). The examples are chosen based on linear scaling of $\vec{\Theta}_0$, which corresponds to the extreme case where $f_3\to 0$ at exactly one point. The dotted line shows the simplest case where $\vec{\Theta}=0$}.
    \label{fig:purity_q3}
\end{figure*}